\documentclass[12pt, draftclsnofoot,onecolumn]{IEEEtran}
\usepackage{dsfont}
\usepackage{amssymb}
\usepackage{subfig}
\usepackage{graphicx, amssymb, amsmath}
\usepackage{extarrows}
\usepackage{algorithm} %format of the algorithm
\usepackage{algorithmic} %format of the algorithm
\usepackage{multirow} %multirow for format of table
\usepackage{color}
\emergencystretch=\maxdimen
\IEEEoverridecommandlockouts
\begin{document}

\title{Cost-Effective Cache Deployment in Mobile Heterogeneous Networks}

\author{Shan~Zhang,~\IEEEmembership{Member,~IEEE,}
	Ning~Zhang,~\IEEEmembership{Member,~IEEE,}
	Peng~Yang,~\IEEEmembership{Student~Member,~IEEE,}
	and~Xuemin~(Sherman)~Shen,~\IEEEmembership{Fellow,~IEEE}% <-this % stops a space
	\thanks{Shan~Zhang, Ning~Zhang and Xuemin~(Sherman)~Shen are with the Department of Electrical and Computer Engineering, University of Waterloo, 200 University Avenue West, Waterloo, Ontario, Canada, N2L 3G1 (email: \{s327zhan, n35zhang, sshen\}@uwaterloo.ca).}% <-this % stops a space
	\thanks{Peng Yang is with the School of Electronic Information and Communications, Huazhong University of Science and Technology, Wuhan, China (email: yangpeng@hust.edu.cn).}% <-this % stops a space
	%\thanks{This work is sponsored in part by the National Basic Research Program of China (973 Program: 2012CB316001), the National Science Foundation of China (NSFC) under grant No. 61201191 and No. 61401250, the Creative Research Groups of NSFC under grant No. 61321061, and Hitachi R\&D Headquarter.}% <-this % stops a space
	%\thanks{Part of this work has been presented in Asilomar Conference on Signals, Systems, and Computers~2015 \cite{mine_asilomar}.}
}

\maketitle

\begin{abstract}
	
	This paper investigates one of the fundamental issues in cache-enabled heterogeneous networks (HetNets): how many cache instances should be deployed at different base stations, in order to provide guaranteed service in a cost-effective manner. Specifically, we consider two-tier HetNets with hierarchical caching, where the most popular files are cached at small cell base stations (SBSs) while the less popular ones are cached at macro base stations (MBSs). For a given network cache deployment budget, the cache sizes for MBSs and SBSs are optimized to maximize network capacity while satisfying the file transmission rate requirements. As cache sizes of MBSs and SBSs affect the traffic load distribution, inter-tier traffic steering is also employed for load balancing. Based on stochastic geometry analysis, the optimal cache sizes for MBSs and SBSs are obtained, which are threshold-based with respect to cache budget in the networks constrained by SBS backhauls. Simulation results are provided to evaluate the proposed schemes and demonstrate the applications in cost-effective network deployment. 
	
\end{abstract}


\section{Introduction}

	Heterogeneous networks (HetNets), consisting of macro base stations (MBSs) and ultra-densely deployed small cell base stations (SBSs), are envisioned as the dominant theme to meet the 1000$\times$ capacity	enhancement in 5G networks and beyond {{ \cite{Ge16_5g_ultra_dense_WC, ultra_dense_SC_2015}.}}
	With network further densified, deploying ideal backhaul with unconstrained capacity for each small cell may be impractical, due the unacceptably high costs of deployment and operation {{ \cite{Yu16_backhaul_cache_GC, Ding12_predoding_backhaul}. }}
	Thus, one of the key problems towards 5G is to reduce the required backhaul capacity while keeping the system capacity. 
	Mobile edge caching provides a promising solution to address the problem, by exploiting the content information \cite{GreenDelivery_SZhou_2015, Yang16_catalyzing_cloud_fog_SDN}.
	As the requested content of mobile users, e.g., video, may show high similarity, caching popular contents at base stations can effectively alleviate the backhaul pressure and enhance network service capability \cite{Qiao_proactive_mmWave}.
	Meanwhile, the delay performance can be significantly improved, with service demands accommodated locally.
	
	Since the study on mobile edge caching is still nascent, many research issues need to be addressed, such as architecture design \cite{Tandon16_fog_mag}, content placement \cite{Li16_cache_min_weighted_load_ICC, Bharath16_learning_caching_TC} and update \cite{Akon13_content_update}.
	However, the caching deployment is overlooked in the existing literature.
	Specifically, the fundamental problem of cache deployment is to optimize the cache sizes of different BSs in HetNets, so as to minimize network deployment and operational costs while guaranteeing quality of service (QoS) performance.
	The basic tradeoff for cache deployment exists between caching efficiency and spectrum efficiency.
	{{On one hand, the contents cached at MBSs can serve more users due to the large cell coverage, providing high caching efficiency.
	On the other hand, the densely deployed SBS tier is more likely to be backhaul-constrained, as extensive spatial spectrum reuse introduces substantial access traffic.
	As a result, deploying more cache instances at SBSs can narrow the gap between backhaul and radio access capacities, and thus improve spectrum efficiency systematically.
	In this regard, cache instances should be deployed appropriately, such that network resources can be balanced and fully utilized \cite{Liu16_EE_cache_JSAC, Liu15_MIMO_cache_TSP}.}}
	However, the cache deployment problem is challenging, as different cache size also influences the traffic load distributions across the network. 
	For example, more traffic needs to be served by MBSs when the MBS cache size increases, changing the loads of both radio access and backhaul of MBS and SBS tiers.
	Therefore, load balancing should be also considered to avoid problems like service outage and resource under-utilization.
	To this end, traffic steering can be leveraged to tune load distribution, and jointly optimized with cache deployment \cite{NZhang16_LTEU}. 
	
	In this paper, the cache deployment problem is investigated in two-tier HetNets, where each SBS caches the most popular files while each MBS caches the less popular ones (i.e., hierarchical caching).
	If cached at the associated MBSs or SBSs, the requested contents will be directly delivered to mobile users through radio access, i.e., content hit.
	Otherwise, the requested contents will be delivered through remote file fetching via backhaul connections, i.e., content miss.
	For a given cache deployment budget, we maximize network capacity while guaranteeing the average file transmission rates, by jointly optimizing the MBS/SBS cache sizes and the inter-tier traffic steering ratio of content miss users.
	However, the problem is of great challenge due to the transmission rate requirements.
	Specifically, file transmission rates depend on both radio and backhaul access conditions, which should account for multi-randomness of traffic load, user location, channel fading and network topology.
	%To address these challenges, we reformulate the problem with decoupled transmission rate constraints on radio and backhaul parts, which can provide the upper bound of network capacity.
	%Additionally, the average transmission rates are derived through stochastic geometry analysis, based on which the cache deployment problem is simplified and numerical results can be obtained.
	Through stochastic geometry analysis, the lower bound of average file transmission rates are derived in closed form, based on which the cache deployment problem is simplified and numerical results can be obtained.
	To offer insights into practical network design, we then focus on the scenario when the MBSs have sufficiently large backhaul capacity while the SBS tier is backhaul constrained.
	The optimal cache deployment is obtained, which is threshold-based with respect to the network cache budget.
	When the cache budget is smaller than certain threshold, all the cache instances should be deployed at SBSs to maximize network capacity.
	When the cache budget exceeds the threshold, the cache deployment problem has multiple optimal solutions to achieve maximal network capacity, and we find the one which can simultaneously maximize content hit rate.
	In fact, cache budget threshold can be interpreted as the deficiency of SBS backhaul, i.e., the minimal cache budget required to match the backhaul and radio resources.
	Moreover, the threshold characterizes the trading relationship between backhaul and cache capacities, which can be applied to cost-effective network deployment.
	
	The contributions of this paper are summarized as follows:
	\begin{enumerate}
		\item The average file transmission rates in large-scale cache-enabled HetNets are analyzed theoretically, considering the constraints of both backhaul capacities and radio resources;
		\item The cache deployment is optimized in HetNets, which maximizes QoS-guaranteed network capacity with the given cache budget;
		\item The inter-tier traffic steering is jointly optimized to balance the loads of MBS and SBS tiers, considering the influence of cache deployment on traffic distributions;
		\item The proposed method can provide the cost-optimal combination of backhaul and radio resource provisioning, which can be applied to practical cache-enabled HetNet deployment.
	\end{enumerate} 
	
	The remaining of this paper is organized as follows. Firstly, related work on mobile edge caching is reviewed in Section~\ref{sec_review}.
	Then, the system model is presented in Section~\ref{sec_system_model}, and the cache deployment problem is formulated in Section~\ref{sec_formulation}. In Section~\ref{sec_capacity_analysis}, the QoS-constrained network capacity is obtained, based on which the optimal cache deployment is analyzed in Section~\ref{sec_solution}. The analytical results are validated through extensive simulations in Section~\ref{sec_simulation}, followed by the cost-effective network deployment illustrations with numerical results. Finally, Section~\ref{sec_conclusions} summarizes the work and discusses future research topics.

%%%%%%%%%%%%%%%%%%%%%%%%%%%%%%%%%%%%%%%%%%%%%%%%%%%%%%%%%%%%%%%%%%%%%%%%%%%%%%%%%%%%%%%%%%%%%%%%%%%
\section{Literature Review}
	\label{sec_review}
	Content caching at mobile edge networks is considered as a promising solution to cope with the mismatch between explosive mobile video traffic and limited backhaul/wireless capacity, which has drawn increasing attention recently.
	Cache-enabled 5G network architectures have been designed in \cite{Wang14_cache_framework_wireless_mag, Bastug14_cache_framework_BS_D2D_mag}, which were shown to have a great potential to reduce mobile traffic through trace-driven simulations.
	The performance of cache-enabled networks has also been analyzed theoretically, which was demonstrated to be more spectrum-efficient compared with the conventional HetNets in backhaul-constrained cases \cite{Liu16_EE_cache_JSAC}.
	{{Meanwhile, effective cache placement schemes have been devised with respect to different optimization objectives, such as maximizing content hit rate \cite{Akon12_cache_BBCR, Serbetci16_multi_PPP_hit_WCNC, Chen16_probabilistic_cache_analysis}, reducing file downloading delay \cite{Chang16_ICC_cache_matching, Cui16_cache_place_delay, Liu16_cache_delay_distributed_algorithm_ICC, bacstuug2016delay}, enhancing user quality of experience (QoE) \cite{Sun_15_cache_QoE_vehicular_streaming}, improving mobility support \cite{Wang16_cache_mobility_mag, Qiao15_video_buffer}, and minimizing specific cost functions \cite{Gregori16_D2D_caching_JSAC, Tao16_TWC_caching_beamforming}.}}

	Although the existing cache placement schemes were designed based on the predefined cache size for each BS, studies on cache deployment were quite limited.
	In the very recent work \cite{Ghoreishi16_cache_multi_layer_video_provisioning_ICC}, the storage costs of different network entities (like remote servers, gateways, and BSs) have been considered, and a multi-layered cache deployment scheme was proposed to maximize the ratio of content hit rate to storage cost.
	The performances of BS-caching and gateway-caching have been compared in \cite{Han16_cache_compare_core_BS}, based on which the cache deployment was optimized to achieve Pareto optimal spectrum efficiency and content hit rate.
	The BS cache sizes are optimized to maximize the minimal user success probability, under the constraints of backhaul capacity and cache deployment budget \cite{Peng16_cache_size_ICC}.
	Insightful as it is, the algorithm in \cite{Peng16_cache_size_ICC} mainly focused on small-scale networks.
	Different from existing work, this paper investigates the cache deployment problem in large-scale HetNets for the first time, aiming at maximizing network capacity while meeting the QoS requirements in terms of file transmission rate.
	Meanwhile, the cache sizes of different BSs are jointly optimized with inter-tier traffic steering.
	The analytical results have taken into account the multi-randomness of network topology, traffic distribution and channel fading, which can provide a guideline for practical network design with mobile edge caching.

%%%%%%%%%%%%%%%%%%%%%%%%%%%%%%%%%%%%%%%%%%%%%%%%%%%%%%%%%%%%%%%%%%%%%%%%%%%%%%%%%%%%%%%%%%%%%%%%%%%
\section{System Model}
    \label{sec_system_model}
    %\input{Model.tex}
    % 4 pages
    % 1 figure of scenario
    % 1 figure of service process: served by which BS, outage
    % 1 notation table
    
    In this section, we present system model and the hierarchical caching framework, with important notations summarized in Table~\ref{tab_notation}.
    
    \subsection{Cache-Enabled Heterogeneous Network Architecture}
    % topology, functions of the two tiers: control, capacity
    % bandwidth, dual connectivity
    % backhaul, cache
    In 5G HetNets, MBSs are responsible for network coverage with control signaling, whereas SBSs are expected to be densely deployed to boost network capacity in a ``plug-and-play'' manner.
    The topology of MBSs are modeled as regular hexagonal cells with density $\rho_\mathrm{m}$, while the distribution of SBSs are modeled as Poisson Point Process (PPP) of density $\rho_\mathrm{s}$.
    MBSs and SBSs use orthogonal spectrum bands to avoid inter-tier interference, and the spectrum reuse factor within each tier is set to be 1.
    Denote by $W_\mathrm{m}$ and $W_\mathrm{s}$ the bandwidths available to each MBS and SBS, respectively.
    Both MBSs and SBSs are connected with core network through wired backhauls, with capacities denoted as $U_\mathrm{MBH}$ and $U_\mathrm{SBH}$, respectively.
    %Meanwhile, both the MBSs and SBSs are cache-enabled to store popular files, so as to reduce backhaul pressure and improve delay performance.
    %Important notations are summarized in Table~\ref{tab_notation}.
    
\begin{figure}[!t]
    	\centering
    	\includegraphics[width=3.2in]{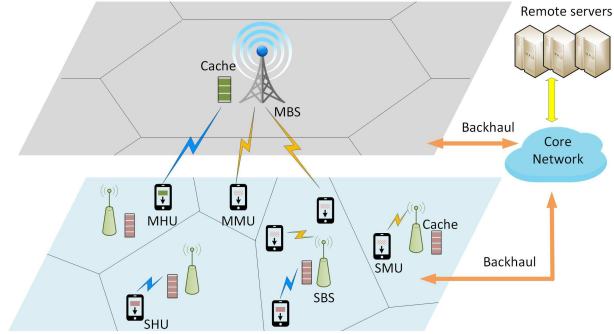}
    	\caption{Cache-enabled heterogeneous networks.}
    	\label{fig_scenario}
    \end{figure}

    The distribution of active users is modeled as a PPP of density $\lambda$, independent of the location of MBSs and SBSs.
    The service process is illustrated as Fig.~\ref{fig_scenario}.
    Each user keeps dual connectivity with an MBS and an SBS \cite{LTE_standard, Ismail13_multihoming_twc, Song12_multi_radio_TWC}, where the MBS (and SBS) which provides the highest average intra-tier signal to interference ratio (SINR) is selected for association.
    The coverage area of each MBS is a hexagonal cell with side length of $D_\mathrm{m}$ ($\frac{3\sqrt{3}}{2} {D_\mathrm{m}}^2 = \frac{1}{\rho_\mathrm{m}}$), and small cells form the Voronoi tessellation, as shown in Fig.~\ref{fig_scenario}.
    As for the service process, mobile users can be directly served through radio access networks if the required files are cached at the MBS or SBS tiers (i.e., content-hit users).
    Instead, content-miss users will randomly choose the associated SBS or MBS with probability $\varphi$ and $1-\varphi$, and the chosen SBS/MBS needs to fetch the required file from remote servers via backhaul.
    Define $\varphi$ as the inter-tier \emph{traffic steering ratio}, which influences the load of MBS and SBS tiers.
    
    \subsection{Hierarchical Caching}
    % file libarary, popularity
    % traffic distribution:PPP
    % association, services: three types of users
    
    \begin{table}[!t]
    	\caption{Important notations}
    	\label{tab_notation}
    	\centering
    	\begin{tabular}{cc|cc}
    		\hline
    		\hline
    		% Notation & Meaning & Notation & Meaning \\
    		% \hline
    		$\rho_\mathrm{m}$ & MBS density & $\rho_\mathrm{s}$ & SBS density \\
    		$W_\mathrm{m}$ & MBS bandwidth & $W_\mathrm{s}$ & SBS bandwidth \\
    		$U_\mathrm{MBH}$ & MBS backhaul capacity & $U_\mathrm{SBH}$ & SBS backhaul capacity \\
    		$F$	& number of files & $q_f$ & popularity of file-$f$ \\
    		$C_\mathrm{m}$ & MBS cache size & $C_\mathrm{s}$ & SBS cache size \\
    		$P_\mathrm{Hit}^{(\mathrm{m})}$ & MBS content hit rate & $P_\mathrm{Hit}^{(\mathrm{s})}$ & SBS content hit rate \\
    		$C$ & network cache budget	& $P_\mathrm{Hit}$ & aggregated content hit rate \\
    		$\lambda$ & traffic density & $\sigma^2$ & noise power\\
    		$\varphi$ & traffic steering ratio & $\xi_c$ & cache deployment ratio \\
    		%$\lambda_\mathrm{MR}$ & traffic density for MBS radio access & $\lambda_\mathrm{MBH}$ & traffic density for MBS backhaul \\
    		%$\lambda_\mathrm{SR}$ & traffic density for SBS radio access & $\lambda_\mathrm{SBH}$ & traffic density for SBS backhaul \\
    		$\check{R}_\mathrm{RAN}$ & required rate at RAN & $\check{R}_\mathrm{BH}$ & required rate at backhaul\\
    		%$I_\mathrm{s}$ & SBS inter-cell interference & $I_\mathrm{m}$ & MBS inter-cell interference \\
    		\hline
    		\hline
    	\end{tabular}
    \end{table}

    Denote by $\mathcal{F}=\left\{1,2,...,f,...F\right\}$ the set of files that may be requested, and denote by $\mathcal{Q} = \left\{ q_1, q_2,...,q_f,...,q_F \right\}$ the popularity distribution ($\sum_{f=1}^{F} q_f=1$, $q_f>0$ for $f=1,2,...,F$).
    Without loss of generality, we assume the files are sorted with descending popularity ($q_f \geq q_{f+1}$ for $f = 1,2,...,F-1$) and have the same size of $L$\footnote{If files have different sizes, they can be divided into the same size to conduct analysis.}.
    A hierarchical content caching framework is adopted, where the SBS tier caches the most popular files while the MBS tier caches the less popular ones to increase content diversity.
    Denote by $C_\mathrm{m}$ and $C_\mathrm{s}$ the cache sizes of each MBS and SBS, respectively.
    %As users are covered simultaneously by both tiers, MBSs and SBSs can cache different files to increase diversity gain.
    Thus, files $\left\{ 1,2,...,C_\mathrm{s} \right\}$ are cached at each SBS, and files $\left\{ C_{\mathrm{s}+1}, C_{\mathrm{s}+2}, ..., C_{\mathrm{s}} + C_{\mathrm{m}} \right\}$ are cached at each MBS.
    Then, the content hit rates of MBS and SBS tiers, $P_\mathrm{Hit}^{(\mathrm{m})}$ and $P_\mathrm{Hit}^{(\mathrm{s})}$, are given by
    \begin{equation}
    	\label{eq_P_hit_s_m}
    	P_\mathrm{Hit}^{(\mathrm{m})} = \sum\limits_{C_\mathrm{s}+1}\limits^{C_\mathrm{m}+C_\mathrm{s}} q_f,~~~~P_\mathrm{Hit}^{(\mathrm{s})} = \sum\limits_{f=1}\limits^{C_\mathrm{s}} q_f,
    \end{equation}
    and total content hit rate is 
    \begin{equation}
    	\label{eq_P_hit}
    	P_\mathrm{Hit} = P_\mathrm{Hit}^{(\mathrm{m})} + P_\mathrm{Hit}^{(\mathrm{s})}= \sum\limits_{f=1}\limits^{C_\mathrm{m}+C_\mathrm{s}} q_f.
    \end{equation}
    % Notice that $C_\mathrm{s}+C_\mathrm{s}$ can be treated as the equivalent cache size for each user.
    %The hierarchical caching has two-fold benefits: (1) maximal content hit rate; (2) high spectrum efficiency.
    With the dual connectivity, the equivalent cache size for each mobile user is $C_\mathrm{m}+C_\mathrm{s}$ according to Eq.~(\ref{eq_P_hit})\footnote{``Eq.'' is short for Equation, and ``Eqn.'' is short for inequation}.
    {{Thus, caching the most popular $C_\mathrm{m}+C_\mathrm{s}$ files can maximize content hit rate, since each mobile user can be served only by the associated MBS or SBS with no intra-tier BS cooperation.
    		In addition, caching more contents at SBSs instead of MBSs can steer more users to the SBS-tier from MBSs.
    		As SBSs are more densely deployed than MBSs in practical networks, steering traffic to the SBS tier can fully utilize rich radio resources with inter-tier load balancing.}}
    
    Define $C$ the network caching budget, i.e., the number of files cached per unit area:
    \begin{equation}
    	\label{eq_cache_size}
    	C = \rho_\mathrm{m} C_\mathrm{m} + \rho_\mathrm{s} C_\mathrm{s}.
    \end{equation}
    Cache deployment determines $C_\mathrm{m}$ and $C_\mathrm{s}$ to optimize network performance, for the given network caching budget $C$.

    \subsection{File Transmission Rate}
    % successful file transmission: data rate demand
    % three types of users, SBS-hit, MBS-hit, SBS-miss: rate analysis
    % outage happens--> file transmission failer, outage probability below certain threshold
    
    With hierarchical caching, users can be classified into four types: (1) MHU (MBS-hit-users), served by the MBS tier with cached contents; (2) SHU (SBS-hit-users), served by the SBS tier with cached contents; (3) MMU (MBS-missed-users), served by the MBS tier through backhaul file fetching; and (4) SMU (SBS-missed-users), served by the SBS tier through backhaul file fetching.
    Based on the properties of PPP, the four types of users also follow independent PPPs, with densities of $P_\mathrm{Hit}^{(\mathrm{m})} \lambda$, $P_\mathrm{Hit}^{(\mathrm{s})} \lambda$, $ (1-P_\mathrm{Hit}) (1-\varphi) \lambda$, and $(1-P_\mathrm{Hit}) \varphi \lambda$, respectively.
    The file transmission rates of MHUs and SHUs only depend on the radio access (i.e., wireless part), whereas the rates of MMUs and SMUs are also constrained by the limited backhaul capacities.
    
    Consider a typical mobile user-$u$.
    If user-$u$ is served by the MBS tier, the achievable rate for radio access is given by
    \begin{equation}
    	\label{eq_R_mr_define}
    	R_\mathrm{MR} = \frac{W_\mathrm{m}}{N_\mathrm{MR}+1} \log_2 \left( 1 + \gamma_\mathrm{m} \right),
    \end{equation}
    where $N_\mathrm{MR}$ denotes the number of residual users being served by the associated MBS except user-$u$ (both MHUs and MMUs included), $\gamma_\mathrm{m}$ is the received SINR given by
    \begin{equation}
    	\label{eq_R_mr_define_1}
    	\gamma_\mathrm{m} = \min \left( \gamma_{\max}, \frac{P_\mathrm{TM} h_\mathrm{m} {d_\mathrm{m}}^{-\alpha_\mathrm{m}}}{\sigma^2 + I_\mathrm{m} } \right),
    \end{equation}
    $\gamma_{\max}$ is the maximal received SINR, $P_\mathrm{TM}$ is the MBS transmit power, $h_\mathrm{m}$ is an exponential random variable with mean 1 incorporating the effect of Rayleigh fading, $\alpha_\mathrm{m}$ is the path loss exponent of the MBS-tier, $d_\mathrm{m}$ denotes the distance from user-$u$ to the associated MBS, $\sigma^2$ is the addictive noise power, and $I_\mathrm{m}$ represents inter-cell interference from other MBSs.
    In practical systems, $N_\mathrm{MR}$ varies randomly with the dynamic arrival and departure of file transmission demands, and $d_\mathrm{m}$ is also uncertain from the network perspective.
    
    If user-$u$ is served by the SBS tier, the achievable rate for radio access can be given by
    \begin{equation}
    	\label{eq_R_sr_define}
    	R_\mathrm{SR} = \frac{W_\mathrm{s}}{N_\mathrm{SR}+1} \log_2 \left( 1 + \gamma_\mathrm{s} \right),
    \end{equation}
    where $N_\mathrm{SR}$ denotes the number of residual users being served by the associated SBS except user-$u$ (including both SHUs and SMUs), 
    \begin{equation}
    	\label{eq_R_sr_define_1}
    	\gamma_\mathrm{s} = \min \left( \gamma_{\max}, \frac{P_\mathrm{TS} h_\mathrm{s} {d_\mathrm{s}}^{-\alpha_\mathrm{s}}}{\sigma^2 + I_\mathrm{s} } \right),
    \end{equation}
    $P_\mathrm{TS}$ is the SBS transmit power, $h_\mathrm{s}$ is an exponential random variable with mean 1 incorporating the effect of Rayleigh fading, $\alpha_\mathrm{s}$ is the path loss exponent of the SBS-tier, $d_\mathrm{s}$ denotes the distance from user-$u$ to the associated SBS, and $I_\mathrm{s}$ represents the inter-cell interference from other SBSs.
    Similarly, $N_\mathrm{SR}$ and $d_\mathrm{s}$ are also random variables. 
    In addition, the probability distribution functions (PDFs) of $N_\mathrm{SR}$ and $d_\mathrm{s}$ can be more complex due to the uncertain small cell sizes.
    
    The MBS and SBS backhaul transmission rates only depend on the corresponding traffic loads and capacities:
    \begin{equation}
    	\label{eq_r_BH}
    	R_\mathrm{MBH} = \frac{U_\mathrm{MBH}}{N_\mathrm{MBH}+1}, ~~~ R_\mathrm{SBH} = \frac{U_\mathrm{SBH}}{N_\mathrm{SBH}+1},
    \end{equation}
    where $N_\mathrm{MBH}$ (or $N_\mathrm{SBH}$) represents the number of residual MMUs (or SMUs) sharing the MBS (or SBS) backhaul expect the considered user-$u$.
    %Denote by $R_\mathrm{MHU}$, $R_\mathrm{SHU}$, $R_\mathrm{MMU}$ and $R_\mathrm{SMU}$ the file transmission rates of a MHU, SHU, MMU and SMU, respectively.
    %Then,
    %	\begin{equation}
    %		\label{eq_rate_relation}
    %		\begin{split}
    %			R_\mathrm{MHU} = R_\mathrm{MR}, ~~ &  R_\mathrm{MMU} = \min\left\{ R_\mathrm{MR}, R_\mathrm{MBH} \right\}, \\
    %			R_\mathrm{SHU} = R_\mathrm{SR}, ~~ &  R_\mathrm{SMU} = \min\left\{ R_\mathrm{SR}, R_\mathrm{SBH} \right\}.
    %		\end{split}
    %	\end{equation}
    %The file transmission rate should to be guaranteed for successful file transmission.
    %In this regard, the average file transmission rates of different users are guaranteed to be no smaller than some threshold as QoS constraints:
    %	\begin{subequations}
    %		\label{eq_rate_define}
    %		\begin{align}
    %			\mathds{E}[ R_\mathrm{MHU} ] & > R_\mathrm{th},\\
    %			\mathds{E}[ R_\mathrm{MMU} ] & > R_\mathrm{th},\\
    %			\mathds{E}[ R_\mathrm{SHU} ] & > R_\mathrm{th},\\
    %			\mathds{E}[ R_\mathrm{SMU} ] & > R_\mathrm{th},
    %		\end{align}
    %	\end{subequations}
    %where $\eta$ is a constant reflecting the strictness of QoS requirements.
    
%%%%%%%%%%%%%%%%%%%%%%%%%%%%%%%%%%%%%%%%%%%%%%%%%%%%%%%%%%%%%%%%%%%%%%%%%%%%%%%%%%%%%%%%%%%%%%%%%%%%
\section{Capacity-Optimal Caching Formulation}
    \label{sec_formulation}
	\begin{figure}[!t]
		\centering
		\includegraphics[width=3in]{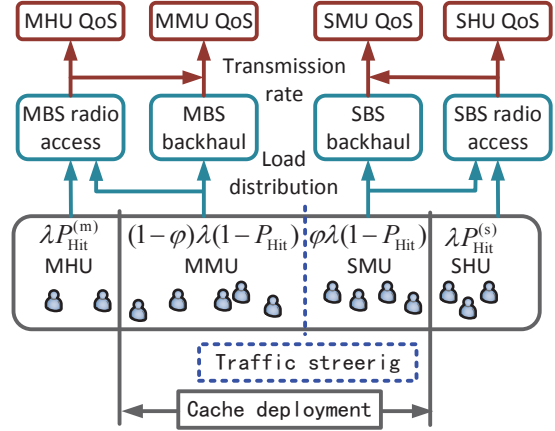}
		\caption{Influence of cache deployment and traffic steering on QoS performance.}
		\label{fig_analytical_framework}
	\end{figure}
	
	To meet QoS requirements, the transmission rates should to be guaranteed for successful file delivery, which depend on the transmission rates of each network part (i.e., MBS radio access, MBS backhaul, SBS radio access and SBS backhaul), as shown in Fig.~\ref{fig_analytical_framework}.
	In addition, cache deployment (i.e., cache sizes $[C_\mathrm{m}, C_\mathrm{s}]$) determines the traffic distributions across the network together with traffic steering ratio $\varphi$, thus influencing the transmission rates.
	%For example, larger MBS cache size increases the content hit rate $P_\mathrm{Hit}^\mathrm{(m)}$, raising the load of MBS tier.
	Therefore, cache deployment should be jointly optimized with traffic steering, which can be formulated as follows:
	\begin{subequations}
		\label{eq_P2}
		\begin{align}
			\max\limits_{C_\mathrm{s},\varphi}~~~& \mu(C_\mathrm{s},\varphi) \\
			(\mbox{P1})~~\mbox{s.t.}~~~~& \mathds{E}\left[ R_\mathrm{MR} \right] \geq \check{R}_\mathrm{RAN},\\
			& \mathds{E}\left[ R_\mathrm{MBH} \right]  \geq \check{R}_\mathrm{BH},\\
			& \mathds{E}\left[ R_\mathrm{SR} \right] \geq \check{R}_\mathrm{RAN},\\
			& \mathds{E}\left[ R_\mathrm{SBH} \right] \geq \check{R}_\mathrm{BH}, \\
			& 0 \leq C_\mathrm{s} \leq C/\rho_\mathrm{s}, ~~ 0 \leq \varphi \leq 1,
		\end{align}
	\end{subequations}
	%\begin{subequations}
	%	\label{eq_P1}
	%	\begin{align}
	%	\max\limits_{C_\mathrm{s},\varphi}~~~& \mu(C_\mathrm{s},\varphi) \\
	%	(\mbox{P1})~~\mbox{s.t.}~~~~& \mathds{E}[ R_\mathrm{MHU} ] \geq R_\mathrm{th}, \mathds{E}[ R_\mathrm{SHU} ] \geq R_\mathrm{th}, \\
	%	& \mathds{E}[ R_\mathrm{MMU} ] \geq R_\mathrm{th}, \mathds{E}[ R_\mathrm{SMU} ] \geq R_\mathrm{th},\\
	%	& 0 \leq C_\mathrm{s} \leq C/\rho_\mathrm{s}, ~~ 0 \leq \varphi \leq 1,
	%	\end{align}
	%\end{subequations}	
	where the objective function $\mu(C_\mathrm{s},\varphi) = \underset{\lambda}{\max}{\left(\lambda |_{\{C_\mathrm{s},\varphi\}}\right)}$ is the network capacity for the given SBS cache size $C_\mathrm{s}$ and traffic steering ratio $\varphi$ (i.e., the maximal traffic density that can be catered), $\check{R}_\mathrm{RAN}$ and $\check{R}_\mathrm{BH}$ denote the per user rate requirements for radio access and backhaul transmissions\footnote{{{In practical systems, the backhaul rate requirement $\check{R}_\mathrm{BH}$ is generally much higher than that of the radio access part $\check{R}_\mathrm{RAN}$, considering the end-to-end delay and the serial transmission structure.}}}, respectively.
	{{The cache size of MBSs $C_\mathrm{m}$ can be determined with $C_\mathrm{s}$ according to Eq.~(\ref{eq_cache_size}).}}
	Constraint (\ref{eq_P2}b) and (\ref{eq_P2}d) guarantee the QoS requirements of content hit users, while (\ref{eq_P2}c) and (\ref{eq_P2}e) further account for the file fetching delay requirements of content miss users.
	{{The average transmission rate is adopted for QoS guarantee as the services suitable for pro-active caching are mostly elastic in practical networks, such as popular video streaming. 
			Furthermore, when the average per user transmission rate is guaranteed, the objective function $\mu(C_\mathrm{s},\varphi)$ can also reflect the network goodput.}}
	
	According to the properties of PPP, the traffic distributions of each network part also follow PPP.
	Denote by $\lambda_\mathrm{MR}$, $\lambda_\mathrm{MBH}$, $\lambda_\mathrm{SR}$ and $\lambda_\mathrm{SBH}$ the equivalent user density for MBS radio access, MBS backhaul, SBS radio access, and SBS backhaul:
	\begin{subequations}
		\label{eq_lambda_define}
		\begin{align}
			\lambda_\mathrm{MR} & = \left[ P_\mathrm{Hit}^{(\mathrm{m})} + (1-P_\mathrm{Hit}) (1-\varphi) \right] \lambda, \\
			\lambda_\mathrm{MBH} & = (1-P_\mathrm{Hit}) (1-\varphi) \lambda, \\
			\lambda_\mathrm{SR} & = \left[ P_\mathrm{Hit}^{(\mathrm{s})} + (1-P_\mathrm{Hit}) \varphi \right] \lambda, \\
			\lambda_\mathrm{SBH} & =  (1-P_\mathrm{Hit}) \varphi \lambda. 
		\end{align}
	\end{subequations}
	The constraints (\ref{eq_P2}b)-(\ref{eq_P2}e) provide the maximal value of $\lambda_\mathrm{MR}$, $\lambda_\mathrm{MBH}$, $\lambda_\mathrm{SR}$, and $\lambda_\mathrm{SBH}$, denoted by $\hat{\lambda}_\mathrm{MR}$, $\hat{\lambda}_\mathrm{MBH}$, $\hat{\lambda}_\mathrm{SR}$, and $\hat{\lambda}_\mathrm{SBH}$, respectively.
	In addition,  $\hat{\lambda}_\mathrm{MR}$, $\hat{\lambda}_\mathrm{MBH}$, $\hat{\lambda}_\mathrm{SR}$, and $\hat{\lambda}_\mathrm{SBH}$ further constrains the traffic arrival rate $\lambda$ with Eq.~(\ref{eq_lambda_define}), for the given $C_\mathrm{s}$ and $\varphi$.
	Thus, the network capacity depends on the bottleneck:
	\begin{equation}
		\begin{split}
			& \mu(C_\mathrm{s},\varphi) = \min\left( \frac{\hat{\lambda}_\mathrm{MR}}{P_\mathrm{Hit}^{(\mathrm{m})} + (1-P_\mathrm{Hit}) (1-\varphi)}, \right.\\
			& \frac{\hat{\lambda}_\mathrm{MBH}}{(1-P_\mathrm{Hit}) (1-\varphi)}, \left.\frac{\hat{\lambda}_\mathrm{SR}}{P_\mathrm{Hit}^{(\mathrm{s})} + (1-P_\mathrm{Hit}) \varphi}, \frac{\hat{\lambda}_\mathrm{SBH}}{(1-P_\mathrm{Hit}) \varphi}\right).			
		\end{split}
	\end{equation}
	%The four constraints of (P1) reflect the service capabilities of wireless and backhaul resources of the MBS-and SBS-tier, and jointly determine network capacity. 
	%In addition, the traffic loads of each part vary with cache deployment and traffic steering.
	%Therefore, the optimal solution of (P1) should match traffic load to service capability at each network part, so as to fully utilize network resources.
	The key issue is the transmission rates analysis, which will be addressed in the next section.

%%%%%%%%%%%%%%%%%%%%%%%%%%%%%%%%%%%%%%%%%%%%%%%%%%%%%%%%%%%%%%%%%%%%%%%%%%%%%%%%%%%%%%%%%%%%%%%%%%%
\section{QoS-Constrained Capacity Analysis}
    \label{sec_capacity_analysis}
    In this section, the file transmission rates of different networks parts are analyzed respectively, based on which the constraints (\ref{eq_P2}b)-(\ref{eq_P2}e) can be simplified with respect to $\lambda$.
    
    \subsection{MBS backhaul}
    
    %	According to Eq.~(\ref{eq_r_BH}), the file transmission rate of a MBS backhaul varies with the number of MMUs served by the corresponding MBS:
    %		 \begin{equation}
    %			 R_\mathrm{MBH} = \frac{U_\mathrm{MBH}}{N_\mathrm{MBH}+1}.
    %		 \end{equation}
    $N_\mathrm{MBH}$ follows Poisson distribution of mean $\lambda_\mathrm{MBH}/\rho_\mathrm{m}$, according to Slivnyak-Mecke theorem \cite{Slivnyak_theorem}.
    Thus, based on Eq.~(\ref{eq_r_BH}), the average file transmission rate of MBS backhaul can be derived:
    \begin{equation}
    \label{eq_rate_MBH}
    \begin{split}
    & \mathds{E} [ R_\mathrm{MBH} ] = \sum\limits_{n=0}^{\infty} \frac{U_\mathrm{MBH}}{n+1} \Pr\left\{ N_\mathrm{MBH} = n \right\} \\
    & = \sum\limits_{n=0}^{\infty} \frac{U_\mathrm{MBH}}{n+1} \frac{\left(\frac{\lambda_\mathrm{MBH}}{\rho_\mathrm{m}}\right)^n}{n!} e^{-\frac{\lambda_\mathrm{MBH}}{\rho_\mathrm{m}}} = \frac{U_\mathrm{MBH} \rho_\mathrm{m}}{\lambda_\mathrm{MBH}} \left( 1-e^{-\frac{\lambda_\mathrm{MBH}}{\rho_\mathrm{m}}} \right).
    \end{split}
    \end{equation}
    Combining Eq.~(\ref{eq_rate_MBH}) with Eq.~(\ref{eq_lambda_define}b), the SBS backhaul constraint Eqn.~(\ref{eq_P2}c) can be simplified with respect to traffic density $\lambda$.
    Denote by $\hat{\lambda}_\mathrm{MBH} = \max\left\{ \lambda_\mathrm{MBH} | \mathds{E} [ R_\mathrm{MBH} ] \geq \check{R}_\mathrm{BH} \right\}$, the maximal traffic load on MBS backhaul.
    As the average rate $\mathds{E} [ R_\mathrm{MBH} ]$ decreases with $\lambda_\mathrm{MBH}$, $\hat{\lambda}_\mathrm{MBH}$ satisfies
    \begin{equation}
    \label{eq_lambda_MBH_max}
    \frac{\rho_\mathrm{m}}{\hat{\lambda}_\mathrm{MBH}} \left( 1-e^{-\frac{\hat{\lambda}_\mathrm{MBH}}{\rho_\mathrm{m}}} \right) = \frac{\check{R}_\mathrm{BH}}{U_\mathrm{MBH}},
    \end{equation}		
    according to Eq.~(\ref{eq_rate_MBH}).
    
    \subsection{SBS backhaul}
    
    Compared with MBS backhaul, the transmission rate of SBS backhaul is more complex due to the random small cell size.
    Denote by $A_\mathrm{s}$ the cell area size, which follows Gamma distribution with shape $\kappa=3.575$ and scale $1/\kappa \rho_\mathrm{s}$ \cite{Cao13_optimal_density_TWC}.
    Thus, the PDF of $A_\mathrm{s}$ is given by
    \begin{equation}
    \label{eq_pdf_A_s}
    f_{A_\mathrm{s}} (A) = A^{\kappa-1} e^{-\kappa \rho_\mathrm{s} A} \frac{(\kappa\rho_\mathrm{s})^\kappa}{\Gamma(\kappa)},
    \end{equation}
    where $\Gamma(\cdot)$ is the gamma function.	
    Furthermore, the number of SMUs served through SBS backhaul follows Poisson distribution of mean $\lambda_\mathrm{SBH} A$ given the cell size $A_\mathrm{s}=A$.
    Thus, based on Eq.~(\ref{eq_r_BH}), the average transmission rate can be derived:
    \begin{equation}
    \label{eq_rate_SBH_derived}
    \begin{split}
    & \mathds{E}[R_\mathrm{SBH}]  = \int\limits_{A=0}\limits^{\infty} \left(\sum_{n=0}^{\infty} \frac{U_\mathrm{SBH}}{n+1}  \Pr\left\{N_\mathrm{SBH}=n | A   \right\}\right) f_{A_\mathrm{s}}(A) \mbox{d} A \\
    & = \int\limits_{A=0}\limits^{\infty} \left( \sum_{n=0}^{\infty} \frac{U_\mathrm{SBH}}{n+1} \frac{(\lambda_\mathrm{SBH}A)^{n}}{n!} e^{-\lambda_\mathrm{SBH}A} \right) f_{A_\mathrm{s}}(A) \mbox{d} A \\
    & = \int\limits_{A=0}\limits^{\infty} \frac{U_\mathrm{SBH}}{\lambda_\mathrm{SBH}A}\left(1-e^{-\lambda_\mathrm{SBH}A}\right) A^{\kappa-1} e^{-\kappa \rho_\mathrm{s} A} \frac{(\kappa\rho_\mathrm{s})^\kappa}{\Gamma(\kappa)} \mbox{d} A \\
    & = \frac{U_\mathrm{SBH}}{\lambda_\mathrm{SBH}} \left\{\int\limits_{A=0}\limits^{\infty} A^{\kappa-2} e^{-\kappa \rho_\mathrm{s} A} \frac{(\kappa\rho_\mathrm{s})^\kappa}{\Gamma(\kappa)} \mbox{d} A \right.\\
    & \left. - \int\limits_{A=0}\limits^{\infty} A^{\kappa-2} e^{-(\kappa \rho_\mathrm{s}+\lambda_\mathrm{SBH}) A} \frac{(\kappa\rho_\mathrm{s})^\kappa}{\Gamma(\kappa)} \mbox{d} A \right\}\\
    & = \frac{U_\mathrm{SBH}\kappa\rho_\mathrm{s}}{\lambda_\mathrm{SBH}} \frac{\Gamma(\kappa-1)}{\Gamma(\kappa)} \left( 1 - \frac{1}{\left(1+\frac{\lambda_\mathrm{SBH}}{\kappa \rho_\mathrm{s}}\right)^{\kappa-1}} \right).
    \end{split}
    \end{equation}
    Combining Eq.~(\ref{eq_rate_SBH_derived}) with Eq.~(\ref{eq_lambda_define}d), the SBS backhaul constraint Eqn.~(\ref{eq_P2}e) can be simplified.	
    According to Eq.~(\ref{eq_rate_SBH_derived}), the maximal traffic load on SBS backhaul $\hat{\lambda}_\mathrm{SBH}$ can be given by
    \begin{equation}
    \label{eq_lambda_SBH_max}
    \frac{\kappa\rho_\mathrm{s}}{\hat{\lambda}_\mathrm{SBH}} \frac{\Gamma(\kappa-1)}{\Gamma(\kappa)} \left( 1 - \frac{1}{\left(1+\frac{\hat{\lambda}_\mathrm{SBH}}{\kappa \rho_\mathrm{s}}\right)^{\kappa-1}} \right) = \frac{\check{R}_\mathrm{BH}}{U_\mathrm{SBH}}.
    \end{equation}
    
    \subsection{MBS Radio Access}
    
    According to Eq.~(\ref{eq_R_mr_define}), the transmission rate of MBS radio access can be given by
    \begin{equation}
    \label{eq_rate_MR_define}
    \begin{split}
    & \mathds{E}[R_\mathrm{MR}] = \underset{ \{ N_\mathrm{MR}, d_\mathrm{m} \}} {\mathds{E}} \left[\frac{W_\mathrm{m}}{1+N_\mathrm{MR}}\log_2(1+\gamma_\mathrm{m})\right] 
    %\sum\limits_{n=0}\limits^{\infty} \int\limits_{0}\limits^{D_\mathrm{m}} \frac{W_\mathrm{m}}{N_\mathrm{MR}} \log_2\left(1+\gamma_\mathrm{m}\right) p_{N_\mathrm{MR}}(n) f_{d_\mathrm{m}} (d) \mbox{d} d,
    \end{split}
    \end{equation}
    where the user number $N_\mathrm{MR}$ follows Poisson distribution of mean $\lambda_\mathrm{MR}/\rho_\mathrm{m}$:	
    \begin{equation}
    \label{eq_p_N_MR}
    p_{N_\mathrm{MR}}(n) = \frac{(\frac{\lambda_\mathrm{MR}}{\rho_\mathrm{m}})^{n}}{n!} e^{-\frac{\lambda_\mathrm{MR}}{\rho_\mathrm{m}}},
    \end{equation}
    and the communication distance $d_\mathrm{m}$ can be considered to follow:
    \begin{equation}
    f_{d_\mathrm{m}}(d) = \frac{2 d}{ {D_\mathrm{m}}^2},
    \end{equation}
    by approximating MBS coverage as a circle of radius $D_\mathrm{m}$.
    Then, the lower bound of average transmission rate for MBS radio access can be obtained by approximating the random inter-cell interference with the average value, which can be quite accurate under the condition of high signal-to-noise ratio (SNR) \cite{mine_TWC_SCoff}.
    
    \textbf{Lemma~1.} The lower bound of average transmission rate of MBS radio access is given by:
    \begin{equation}	
    \label{e_R_MR_aver_lemma_1}
    \mathds{E} [R_\mathrm{MR}] \geq \frac{{\tau}_\mathrm{m} W_\mathrm{m}}{\bar{N}_\mathrm{MR}},
    \end{equation}
    where 
    \begin{equation}
    \begin{split}
    & {\tau}_\mathrm{m} = \log_2\frac{P_\mathrm{TM} D_\mathrm{m}^{-\alpha_\mathrm{m}} }{(1+\theta_\mathrm{m})\sigma^2}+\frac{\alpha_\mathrm{m}}{2 \ln 2}\left(1-\frac{D_\mathrm{min}^2}{D_\mathrm{m}^2}\right), \\
    & \bar{N}_\mathrm{MR} = \frac{\lambda_\mathrm{MR}} {\rho_\mathrm{m}} \left(1-e^{-\frac{\lambda_\mathrm{MR}}{\rho_\mathrm{m}}}\right)^{-1},
    \end{split}
    \end{equation}
    $D_\mathrm{min}$ is the transmission distance corresponding to the maximal received SINR (i.e., $\frac{P_\mathrm{TM}D_\mathrm{min}^{-\alpha_\mathrm{m}}}{(1+\theta_\mathrm{m})\sigma^2} = \gamma_\mathrm{max}$), and $\theta_\mathrm{m}$ denotes the ratio of average inter-cell interference to noise (i.e., $\theta_\mathrm{m} \sigma^2 = \mathds{E}[I_\mathrm{m}] $).
    The equality of Eqn.~(\ref{e_R_MR_aver_lemma_1}) holds when $\frac{\sigma^2}{P_\mathrm{TM}} \rightarrow 0$.
    
    \emph{Proof:}~Please refer to Appendix~\ref{appendix_rate_MR}.
    \hfill \rule{4pt}{8pt}
    
    \emph{Remark:}~The physical meaning of ${\tau}_\mathrm{m}$ is the average spectrum efficiency of MBS tier, and $\bar{N}_\mathrm{MR}$ reflects the average number of users accessing each MBS. In practical cellular networks, the received SINR is usually guaranteed to be high enough for reliable communications, through methods like inter-cell interference control. Therefore, Lemma~1 can be applied to approximate average data rate, and constraint Eqn.~(\ref{eq_P2}b) can be simplified with respect to traffic load $\lambda$ based on Eq.~(\ref{eq_lambda_define}a).	
    In addition, the maximal traffic load on MBS radio access $\hat{\lambda}_\mathrm{MR}$ can be given by:
    \begin{equation}
    \label{eq_lambda_MR_max}
    \frac{\rho_\mathrm{m}}{\hat{\lambda}_\mathrm{MR}}\left(1-e^{-\frac{\hat{\lambda}_\mathrm{MR}}{\rho_\mathrm{m}}}\right) = \frac{\check{R}_\mathrm{RAN}}{{\tau}_\mathrm{m}W_\mathrm{m}}.
    \end{equation}

    \subsection{SBS Radio Access}
    
    According to Eq.~(\ref{eq_R_sr_define}), the average transmission rate for SBS radio access is given by
    \begin{equation}
    \label{eq_rate_MR_define}
    \begin{split}
    & \mathds{E}[R_\mathrm{SR}] = \underset{ \{ A_\mathrm{s}, N_\mathrm{SR}, d_\mathrm{s} \}} {\mathds{E}} \left[\frac{W_\mathrm{s}}{1+N_\mathrm{SR}}\log_2(1+\gamma_\mathrm{s})\right].
    % = \int\limits_{0}\limits^{\infty} \int\limits_{0}\limits^{\infty}\sum\limits_{n=0}\limits^{\infty} \frac{W_\mathrm{s}}{N_\mathrm{SR}} \log_2\left(1+\gamma_\mathrm{s}\right) p_{N_\mathrm{SR}}(n) f_{d_\mathrm{s}} (d) f_{A_\mathrm{s}} (A) \mbox{d} d \mbox{d} A 
    \end{split}
    \end{equation}
    %	where $f_{d_\mathrm{s}}(d)$ and $p_{N_\mathrm{SR}}(n)$ are the PDFs of communication distance and user number at the SBS-tier, for the given cell size $A_\mathrm{s}=A$. 
    %	As the distribution of SBS users follows PPP, we have
    %		\begin{equation}
    %			p_{N_\mathrm{SR}}(n) = \frac{(\lambda_\mathrm{SR} A)^n}{n!} e^{-\lambda_\mathrm{SR} A}.
    %		\end{equation}
    The accurate average transmission rate cannot be derived in closed form, due to the random SBS topology and user location.
    Similarly, the lower bound of average transmission rate can be obtained by approximating the random inter-cell interference by the average value, given by Lemma~2.
    
    \textbf{Lemma~2.}~The lower bound of average transmission rate of SBS radio access is given by:
    \begin{equation}
    \label{e_R_SR_aver_lemma_2}
    \mathds{E} [R_\mathrm{SR}] \geq \frac{\tau_\mathrm{s}W_\mathrm{s}}{\bar{N}_\mathrm{SR}},
    \end{equation} 	
    where
    \begin{equation}
    \begin{split}
    & \tau_\mathrm{s} = \log_2 \frac{P_\mathrm{TS} (\pi \rho_\mathrm{s})^{\frac{\alpha_\mathrm{s}}{2}}}{(1+\theta_\mathrm{s})\sigma^2} + \frac{\alpha_\mathrm{s}}{2 \ln 2} \gamma,\\
    & \bar{N}_\mathrm{SR} = \frac{\lambda_\mathrm{SR}}{\kappa \rho_\mathrm{s}} \frac{\Gamma(\kappa)}{\Gamma(\kappa-1)} \left[1-\frac{1}{\left(1+\frac{\lambda_\mathrm{SR}}{\kappa \rho_\mathrm{s}}\right)^{\kappa-1}}\right]^{-1},
    \end{split}
    \end{equation}
    $\gamma \approx 0.577$ is Euler-Mascheroni constant, and $\theta_\mathrm{s}$ denotes the ratio of average inter-cell interference to noise at SBS tier.
    The equality of Eqn.~(\ref{e_R_SR_aver_lemma_2}) holds when $\frac{\sigma^2}{P_\mathrm{TS}} \rightarrow 0$.
    
    \emph{Proof}: Please refer to Appendix~\ref{appendix_rate_SR}.
    \hfill \rule{4pt}{8pt}
    
    \emph{Remark:}~${\tau}_\mathrm{s}$ can be interpreted as the average spectrum efficiency of SBS tier, and $\bar{N}_\mathrm{SR}$ reflects the average number of users accessing each SBS.
    The constraint Eqn.~(\ref{eq_P2}d) can be simplified by combing Lemma~2 with Eq.~(\ref{eq_lambda_define}c).
    In addition, the maximal traffic load on SBS radio access $\hat{\lambda}_\mathrm{SR}$ can be given by
    \begin{equation}
    \label{eq_lambda_SR_max}
    \frac{\kappa \rho_\mathrm{s}}{\hat{\lambda}_\mathrm{SR}} \frac{\Gamma(\kappa-1)}{\Gamma(\kappa)} \left[1-\frac{1}{\left(1+\frac{\hat{\lambda}_\mathrm{SR}}{\kappa \rho_\mathrm{s}}\right)^{\kappa-1}}\right] = \frac{\check{R}_\mathrm{RAN}}{\tau_\mathrm{s}W_\mathrm{s}} ,
    \end{equation}

%%%%%%%%%%%%%%%%%%%%%%%%%%%%%%%%%%%%%%%%%%%%%%%%%%%%%%%%%%%%%%%%%%%%%%%%%%%%%%%%%%%%%%%%%%%%%%%%%%%
\section{Capacity-Optimal Hierarchical Caching}
	\label{sec_solution}
	Based on the transmission rates analysis, problem (P1) can be simplified as follows:
	\begin{subequations}
		\label{eq_P3}
		\begin{align}
		\max\limits_{C_\mathrm{s},\varphi}~~~& \mu(C_\mathrm{s},\varphi) \\
		(\mbox{P2})~~\mbox{s.t.}~~~~& \left[P_\mathrm{Hit}^{(\mathrm{m})}+(1-P_\mathrm{Hit})(1-\varphi)\right] \lambda \leq \hat{\lambda}_\mathrm{MR},\\
		& (1-P_\mathrm{Hit})(1-\varphi) \lambda \leq \hat{\lambda}_\mathrm{MBH},\\
		& \left[P_\mathrm{Hit}^{(\mathrm{s})}+(1-P_\mathrm{Hit})\varphi\right] \lambda \leq \hat{\lambda}_\mathrm{SR},\\
		& (1-P_\mathrm{Hit})\varphi \lambda \leq \hat{\lambda}_\mathrm{SBH},\\
		& 0 \leq C_\mathrm{s} \leq C/\rho_\mathrm{s}, ~~ 0 \leq \varphi \leq 1,
		\end{align}
	\end{subequations}
	where $\hat{\lambda}_\mathrm{MR}$, $\hat{\lambda}_\mathrm{MBH}$, $\hat{\lambda}_\mathrm{SR}$, and $\hat{\lambda}_\mathrm{SBH}$ are given by Eqs.~(\ref{eq_lambda_MR_max}, \ref{eq_lambda_MBH_max}, \ref{eq_lambda_SR_max}, and \ref{eq_lambda_SBH_max}), while the content hit rate $P_\mathrm{Hit}^{(\mathrm{m})}$, $P_\mathrm{Hit}^{(\mathrm{s})}$ and $P_\mathrm{Hit}$ can be derived by Eqs.~(\ref{eq_P_hit_s_m}) and (\ref{eq_P_hit}) with respect to different caching deployment $[C_\mathrm{s},C_\mathrm{m}]$.
	%%%% different with conventional load balancing
	Although different cache deployments can influence the traffic load distribution, problem (P2) differs significantly from the conventional load balancing problems.
	The total traffic load remains constant in load balancing problems, where the traffic load is shifted from one part to another.
	Instead, different cache deployments may change backhaul loads, as the content hit rate varies with cache sizes.
	
	\subsection{Problem Analysis and Solutions}
	
	Denote by
	\begin{subequations}
		\label{eq_capacity_part}
		\begin{align}
		\mu_\mathrm{MR} & = \frac{\hat{\lambda}_\mathrm{MR}}{ P_\mathrm{Hit}^{(\mathrm{m})} + (1-P_\mathrm{Hit}) (1 - \varphi) }, \\
		\mu_\mathrm{MBH} & = \frac{\hat{\lambda}_\mathrm{MBH}}{(1-P_\mathrm{Hit})(1-\varphi)}, \\
		\mu_\mathrm{SR} & = \frac{\hat{\lambda}_\mathrm{SR}}{ P_\mathrm{Hit}^{(\mathrm{s})} + (1-P_\mathrm{Hit}) \varphi}, \\
		\mu_\mathrm{SBH} & = \frac{\hat{\lambda}_\mathrm{SBH}}{ (1-P_\mathrm{Hit})\varphi },
		\end{align}
	\end{subequations}
	the maximal network traffic density constrained by the corresponding network part, and the network capacity is given by
	\begin{equation}
	\label{eq_P3_opt_simple}
	\mu(C_\mathrm{s},\varphi) = \min\left\{ \mu_\mathrm{MR}, \mu_\mathrm{MBH}, \mu_\mathrm{SR}, \mu_\mathrm{SBH} \right\}.
	\end{equation}	
	Numerical results of problem (P2) can be obtained by exhaustive search based on Eqs.~(\ref{eq_lambda_MBH_max}, \ref{eq_lambda_SBH_max}, \ref{eq_lambda_MR_max}, \ref{eq_lambda_SR_max}, \ref{eq_capacity_part} and \ref{eq_P3_opt_simple}).
	Furthermore, low-complexity heuristic algorithms can be designed.	
	To maximize network capacity, the traffic loads of each network part should be balanced according to the corresponding service capabilities.
	Notice that the traffic load distribution can be manipulated by adjusting cache deployment strategy or traffic steering ratio.
	Specifically, Table~\ref{tab_variation} gives the variations of traffic load distribution with respect to cache size and traffic steering ratio, with proof provided in Appendix~\ref{appendix_variation}.
	Table~\ref{tab_variation} can provide a guideline to enhance network capacity in practical networks.
	For example, when the MBS radio access is the performance bottleneck (i.e., $\mu_\mathrm{MR}<\mu_\mathrm{MBH}$, $\mu_\mathrm{MR}<\mu_\mathrm{SR}$, $\mu_\mathrm{MR}<\mu_\mathrm{SBH}$), we can either reduce cache size at MBSs, or increase the traffic steering ratio.
	Instead, when the SBS backhaul is limited, we can either reduce the SBS cache size or lower the traffic steering ratio.
	
	\begin{table}[!t]
		\caption{Capacity variations by increasing cache size or steering ratio}
		\label{tab_variation}
		\centering
		\begin{tabular}{ccccc}
			\hline
			\hline
			& $\mu_\mathrm{MR}$ & $\mu_\mathrm{MBH}$ & $\mu_\mathrm{SR}$ & $\mu_\mathrm{SBH}$ \\
			\hline
			$C_\mathrm{s}$ & Increase & Decrease & Decrease & Decrease \\
			$\varphi$ & Increase & Increase  & Decrease & Decrease \\
			\hline
			\hline
		\end{tabular}
	\end{table}
	
	\subsection{SBS-Backhaul-Constrained HetNets}
	
	In practical systems, MBSs are expected to be equipped with optical fiber backhauls which can provide sufficiently large bandwidth, whereas the capacity for radio access can be much smaller due to spectrum resource scarcity.
	In this case, problem (P2) can be further simplified by removing constraint (\ref{eq_P3}c).
	The condition of ideal MBS backhaul is $\hat{\lambda}_\mathrm{MR} \leq \hat{\lambda}_\mathrm{MBH}$, whereby constraint (\ref{eq_P3}c) always holds if (\ref{eq_P3}b) is satisfied.	
	
	%% unconstrained backhaul
	In what follows, we focus on HetNets with ideal MBS backhaul, and find the analytical solutions to problem (P2) with different network settings.
	To begin with, we consider a simple case when both the MBS and SBS tiers have unconstrained backhaul capacity, i.e., $\hat{\lambda}_\mathrm{MR} \leq \hat{\lambda}_\mathrm{MBH}$ and $\hat{\lambda}_\mathrm{SR} \leq \hat{\lambda}_\mathrm{SBH}$.
	In this case, constraints (\ref{eq_P3}c) and (\ref{eq_P3}e) can be both neglected, and the network capacity cannot be improved by deploying cache.
	This case corresponds to conventional network deployment, where the backhaul capacity is sufficiently reserved while radio resources are limited.	
	Problem (P2) degenerates to the conventional inter-tier load balancing problem which can be easily solved.
	By adding constraints (\ref{eq_P3}b) and (\ref{eq_P3}d), we have $\lambda \leq \hat{\lambda}_\mathrm{MR} + \hat{\lambda}_\mathrm{SR}$.
	Thus, the maximal network capacity is $\mu(C_\mathrm{s},\varphi)^* = \hat{\lambda}_\mathrm{MR} + \hat{\lambda}_\mathrm{SR}$, and the optimal traffic steering is given by $\varphi^* = \frac{\hat{\lambda}_\mathrm{SR}}{\hat{\lambda}_\mathrm{MR} + \hat{\lambda}_\mathrm{SR}}$ without mobile edge caching.
	%%%%%%%%%%%%%%%%%%%%%%%%%%%%%%
	%%%%%%%%%%%%%%%%%%%%%%%%%%%%%%	
	As SBSs further densifies, the network capacity for radio access can scale almost linearly with SBS density, whereas densely deploying high speed fiber backhaul for each SBS is not practical considering the high cost.
	In addition, the traffic of multiple SBSs can be geographically aggregated and transmitted through a shared backhaul (e.g., the cloud-RAN architecture), which further limits the backhaul capacity of each SBS \cite{CRAN_white_paper}.
	Thus, the SBS tier can be backhaul-constrained, i.e., $\hat{\lambda}_\mathrm{SR} > \hat{\lambda}_\mathrm{SBH}$.
	In this case, deploying caching at SBSs can improve network capacity by reducing backhaul traffic load, which is equivalent to increasing backhaul capacity.
	Furthermore, the optimal solutions to (P2) is threshold-based.
	Denote by $C_\mathrm{min}$ and $C_\mathrm{max}$ the two thresholds of cache budgets, which are given by
	\begin{equation}
	\label{eq_C_min_max}
	\begin{split}
	\sum\limits_{f=1}\limits^{C_\mathrm{min}/\rho_\mathrm{s}} q_f = \frac{\hat{\lambda}_\mathrm{SR}-\hat{\lambda}_\mathrm{SBH}}{\hat{\lambda}_\mathrm{MR}+ \hat{\lambda}_\mathrm{SR}}, & \\
	\sum\limits_{f=C_\mathrm{min}/\rho_\mathrm{s}+1}\limits^{C_\mathrm{min}/\rho_\mathrm{s} + (C_\mathrm{max}-C_\mathrm{min}) / \rho_\mathrm{m} } q_f =  & \frac{\hat{\lambda}_\mathrm{MR} }{ \hat{\lambda}_\mathrm{MR} + \hat{\lambda}_\mathrm{SR} } .
	\end{split}		
	\end{equation}	
	The optimal solution to (P2) is summarized in Propositions~2-4, under different cache budgets.
	
	%% \emph{2) Insufficient Caching Budget}
	\textbf{Proposition~2.} If $C < C_\mathrm{min} $, the optimal solution to (P2) is given by $C_\mathrm{s} = C/ \rho_\mathrm{s}$, $\varphi = \hat{\lambda}_\mathrm{SBH}/(\hat{\lambda}_\mathrm{MR} + \hat{\lambda}_\mathrm{SBH})$.
	
	\emph{Proof}: Please refer to Appendix~\ref{appendix_insufficient_cache}.
	\hfill \rule{4pt}{8pt}
	
	\emph{Remark:} As deploying caching is equivalent to increasing backhaul capacity, cache instances need to be deployed at the backhaul-constrained SBSs for compensation.
	$C_\mathrm{min}$ can be interpreted as the deficiency of SBS backhaul, and $C_\mathrm{min}/\rho_\mathrm{min}$ is the minimal SBS cache size needed to match with radio access resources.	
	When the cache budget is smaller than $C_\mathrm{min}$, the SBS tier is still backhaul-constrained even when all the cache instances are deployed at SBSs, and the network capacity increases with the cache budget.
	Furthermore, the SBS radio resources are always redundant compared with SBS backhaul, and thus the performance bottleneck exists at either the SBS backhaul or the MBS radio access.
	Accordingly, the load of SBS backhaul and MBS radio access should be balanced, by steering the content miss users to the two tiers appropriately.
	
	%%% \emph{3) Sufficient Caching Budget}
	
	When the cache budget increases to $C_\mathrm{min}$, the SBS backhaul deficiency can be completely compensated, and the network capacity achieves $\hat{\lambda}_\mathrm{MR}+\hat{\lambda}_\mathrm{SR}$.
	As cache budget further increases, the network performance will be constrained by radio access instead of SBS backhaul, and the network capacity no longer increases.	
	If $C > C_\mathrm{min}$, there exist solutions to achieving the maximal network capacity $\mu(C_\mathrm{s},\varphi)^* = \hat{\lambda}_\mathrm{MR}+\hat{\lambda}_\mathrm{SR}$, as long as the SBS cache size is large enough to compensate backhaul deficiency, i.e., $C_\mathrm{s} \geq C_\mathrm{min}/\rho_\mathrm{s}$.
	Among these capacity-optimal solutions, those with higher content hit rates can further improve user experience by reducing content fetching delay.
	Thus, we aim to find the solution $\left[C_\mathrm{s}^*, \varphi^*\right]$, which can maximize content hit rate while guaranteeing network capacity $\hat{\lambda}_\mathrm{MR}+\hat{\lambda}_\mathrm{SR}$.
	Although increasing the MBS cache size can improve content hit rate, larger MBS cache size results in heaver traffic load at MBSs, which can degrade the transmission rate at MBS radio access, especially when the cache budget is large.
	Thus, the optimal cache deployment and traffic steering depend on the cache budget, given by Proposition~3 and 4.	
	
	\textbf{Proposition~3.} If $ C_\mathrm{min} \leq C < C_\mathrm{max}$, the solution $[C_\mathrm{s}^*, \varphi^*]$ satisfying
	\begin{subequations}
		\label{eq_P3_opt_sufficient_C}
		\begin{align}
		C_\mathrm{s}^* & = C_\mathrm{min}/\rho_\mathrm{s}, \\
		(1-P_\mathrm{Hit}^*) \varphi^* & = \sum\limits_{f=C_\mathrm{s}^*+C_\mathrm{m}^*+1}\limits^{F} q_f \varphi^* = \frac{\hat{\lambda}_\mathrm{SBH}}{\hat{\lambda}_\mathrm{MR} +\hat{\lambda}_\mathrm{SR}} ,
		% & {P_\mathrm{Hit}^{(\mathrm{s})}}^* = \sum\limits_{f=1}\limits^{C_\mathrm{s}^*} q_f = \frac{\hat{\lambda}_\mathrm{SR} - \hat{\lambda}_\mathrm{SBH} }{\hat{\lambda}_\mathrm{MR} +\hat{\lambda}_\mathrm{SR}},
		\end{align}
	\end{subequations}
	can maximize content hit rate while maximizing network capacity, where $C_\mathrm{m}^*=(C - \rho_\mathrm{s} C_\mathrm{s}^*) / \rho_\mathrm{m}$ and $P_\mathrm{Hit}^*$ is the corresponding aggregated content hit rate.
	
	\emph{Proof}: Please refer to Appendix~\ref{appendix_sufficient_cache}.
	\hfill \rule{4pt}{8pt}

	\textbf{Proposition~4.} If $ C \geq C_\mathrm{max} $, the solution $\left[C_\mathrm{s}^*, \varphi^* \right]$ satisfying
	\begin{subequations}
		\label{eq_P3_opt_oversupplied_C}
		\begin{align}
		\varphi^* & = 1 \\
		{P_\mathrm{Hit}^{(\mathrm{m})}}^* =  \sum\limits_{f=C_\mathrm{s}^*+1}\limits^{C_\mathrm{s}^*+C_\mathrm{m}^*} & = \frac{\hat{\lambda}_\mathrm{MR} }{\hat{\lambda}_\mathrm{MR} +\hat{\lambda}_\mathrm{SR}},
		\end{align}
	\end{subequations}
	can maximize content hit rate while maximizing network capacity, where $C_\mathrm{m}^*=(C - \rho_\mathrm{s} C_\mathrm{s}^*) / \rho_\mathrm{m}$ and $P_\mathrm{Hit}^{(\mathrm{m})}$ is the corresponding content hit rate at MBSs.
	
	\emph{Proof}: Substituting Eq.~(\ref{eq_P3_opt_oversupplied_C}) into (\ref{eq_capacity_part}) and (\ref{eq_P3_opt_simple}), the network capacity can achieve $\hat{\lambda}_\mathrm{MR}+\hat{\lambda}_\mathrm{SR}$. As $\varphi^*=1$, $\mu_\mathrm{MR}$ will decrease as $C_\mathrm{m}$ increases, degrading network capacity. Therefore, $\left[C_\mathrm{s}^*, \varphi^* \right]$ achieves maximal content hit rate among the capacity-optimal schemes.
	\hfill \rule{4pt}{8pt}\\
	
	Based on the above analysis, we summarize the propose capacity-optimal cache deployment scheme for SBS-backhaul-constrained HetNets as follows:
	\begin{itemize}
		\item \textbf{Case-1:} If $C \leq C_\mathrm{min} $, all cache instances should be deployed at the SBS tier;
		\item \textbf{Case-2:} If $C_\mathrm{min} < C \leq C_\mathrm{max} $, the cache size of each SBS is $C_\mathrm{min}/\rho_\mathrm{s}$, and the remaining cache budget should be deployed at the MBS tier (i.e., $C_\mathrm{m} = (C- C_\mathrm{min})/\rho_\mathrm{m}$);
		\item \textbf{Case-3:} If $C > C_\mathrm{max} $, the optimal cache deployment should guarantee that MBS-tier content hit rate satisfies Eq.~(\ref{eq_P3_opt_oversupplied_C}b).
	\end{itemize}
	Meanwhile, traffic steering ratio should be adjusted with caching deployment, to balance inter-tier traffic load.
	%	The proposed cache deployment scheme can achieve high network capacity and content hit rate.
	In addition, the analytical results of thresholds $C_\mathrm{min}$ and $C_\mathrm{max}$ are derived as Eq.~(\ref{eq_C_min_max}), which depend on backhaul and radio resource provisions.

%%%%%%%%%%%%%%%%%%%%%%%%%%%%%%%%%%%%%%%%%%%%%%%%%%%%%%%%%%%%%%%%%%%%%%%%%%%%%%%%%%%%%%%%%%%%%%%%%%%
\section{Simulation and Numerical Results}
    \label{sec_simulation}
    In this section, simulations are conducted to validate the obtained analytical results, and numerical results are provided to offer insights into practical network deployment.
    The file popularity is considered to follow Zipf distribution \cite{video_popularity_2009}:
    \begin{equation}
    q_f = \frac{1/f^\nu}{\sum_{h=1}^{F} 1/h^\nu},
    \end{equation}
    where $\nu\geq 0$ indicates the skewness of popularity distribution. %For example, a larger $\nu$ corresponds to more concentrated file requests.
    In this simulation, $\nu$ is set as $0.56$, featuring video streaming services \cite{video_popularity_2009}.
    Important parameters are listed in Table~\ref{tab_parameter} \cite{Liu16_EE_cache_JSAC}.
    
    \begin{table}[!t]
    	\caption{Simulation parameters}
    	\label{tab_parameter}
    	\centering
    	\begin{tabular}{cccc}
    		\hline
    		\hline
    		Parameter & Value & Parameter & Value \\
    		\hline
    		$D_\mathrm{m}$ & 500 m & $\rho_\mathrm{s}$ & 50 /km$^2$ \\
    		$P_\mathrm{TM}$ & 10 W & $P_\mathrm{TS}$ & 2 W  \\
    		$W_\mathrm{M}$ & 100 MHz & $W_\mathrm{S}$ & 10 MHz \\
    		$\alpha_\mathrm{m}$ & 3.5 & $\alpha_\mathrm{s}$ & 4 \\
    		$\sigma^2$ & -105 dBm/MHz & $U_\mathrm{MBH}$ & 100 Gbps\\
    		$\theta_\mathrm{m}$ & 1000 & $\theta_\mathrm{s}$ & 1000\\
    		$\check{R}_\mathrm{RAN}$ & 5 Mbps & $\check{R}_\mathrm{BH}$ & 50 Mbps \\
    		$F$ & 1000 & $\nu$ & 0.56 \\		
    		%$U_\mathrm{MBH}$ & 20 Mbps & $U_\mathrm{SBH}$ & 1 Mbps \\
    		\hline
    		\hline
    	\end{tabular}
    \end{table}

    \subsection{Analytical Results Evaluation}
    
\begin{figure}[!t]
    	\centering
    	\includegraphics[width=2.5in]{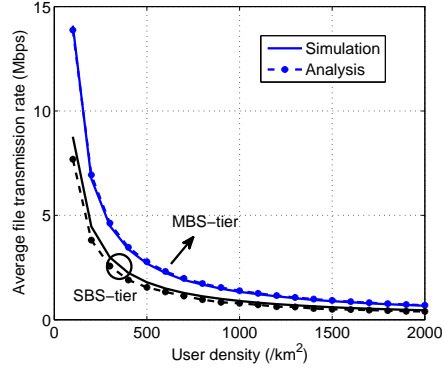}
    	\caption{{{Evaluation of derived file tranmsission rate at radio access parts.}}}
    	\label{fig_evaluation}
    \end{figure}

    {{The analytical results of file transmission rates for MBS/SBS radio access are validated in Fig.~\ref{fig_evaluation}, where 15\% traffic is served by the MBS tier and the remaining is steered to the SBS tier.}}
    Monte Carlo method is applied in simulation, with SBS topology, user location and channel fading generated according to the corresponding PDFs.
    The simulation results is averaged over 10000 samples.
    The analytical results are calculated based on Lemmas~1 and 2.
    As the analytical and simulation results are shown to be very close, Lemmas~1 and 2 can be applied to approximate transmission rate analysis for radio access.	
    
    \begin{figure*}[!t]
    	\centering
    	\subfloat[$C$=50 files/km$^2$] {\includegraphics[width=2.2in]{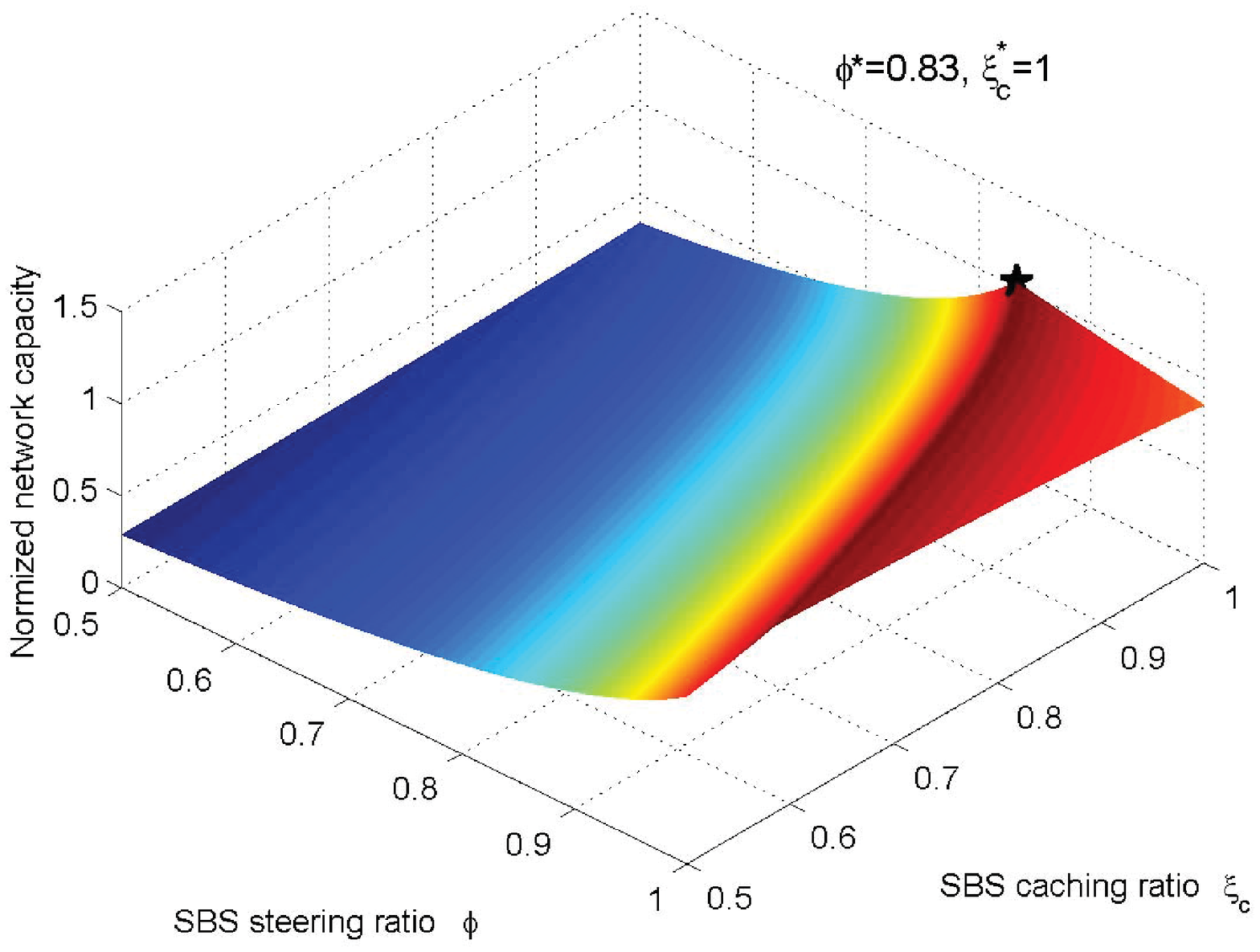}}
    	%\hfil
    	\subfloat[$C$=900 files/km$^2$]{\includegraphics[width=2.2in]{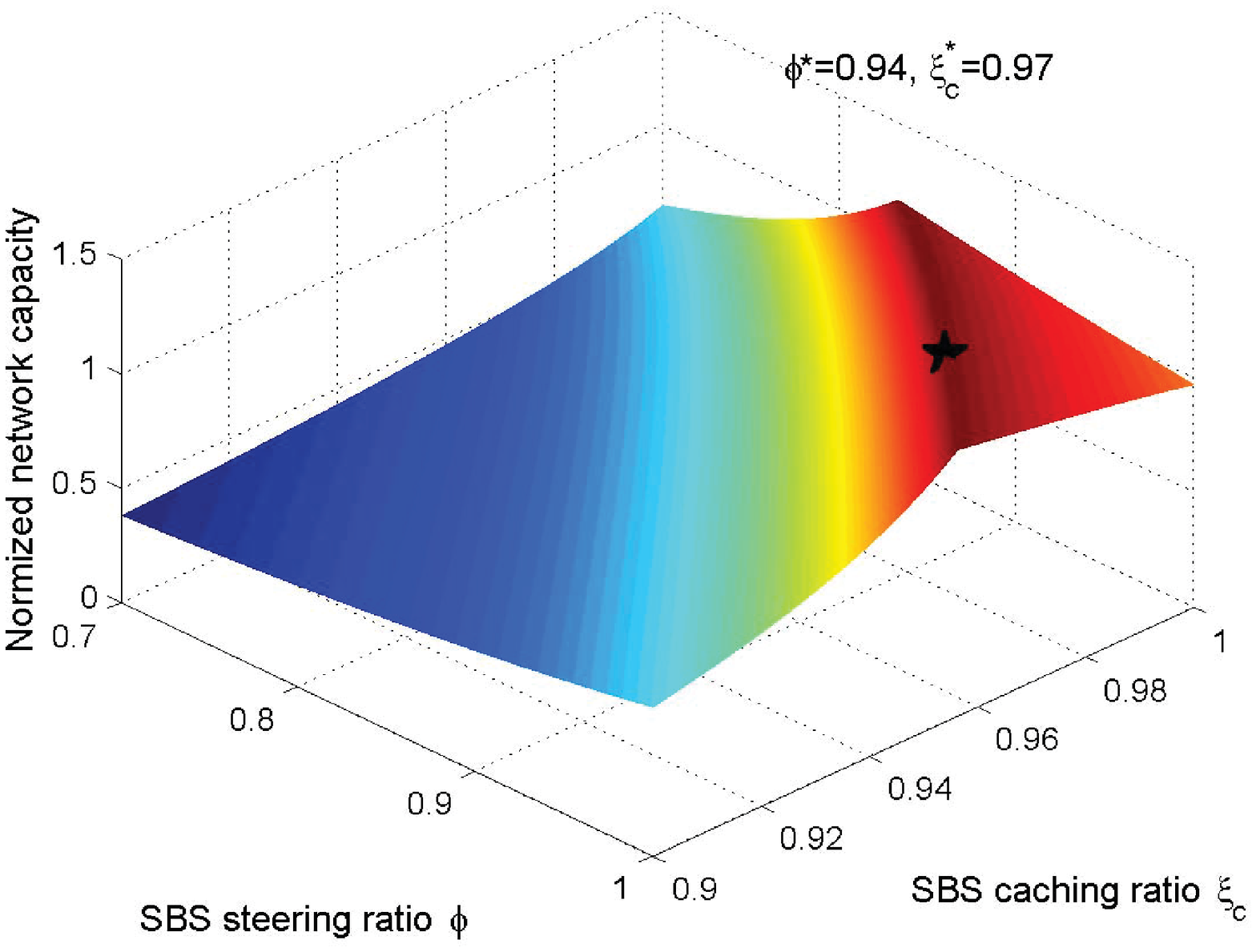}}
    	%\hfil
    	\subfloat[$C$=1000 files/km$^2$]{\includegraphics[width=2.2in]{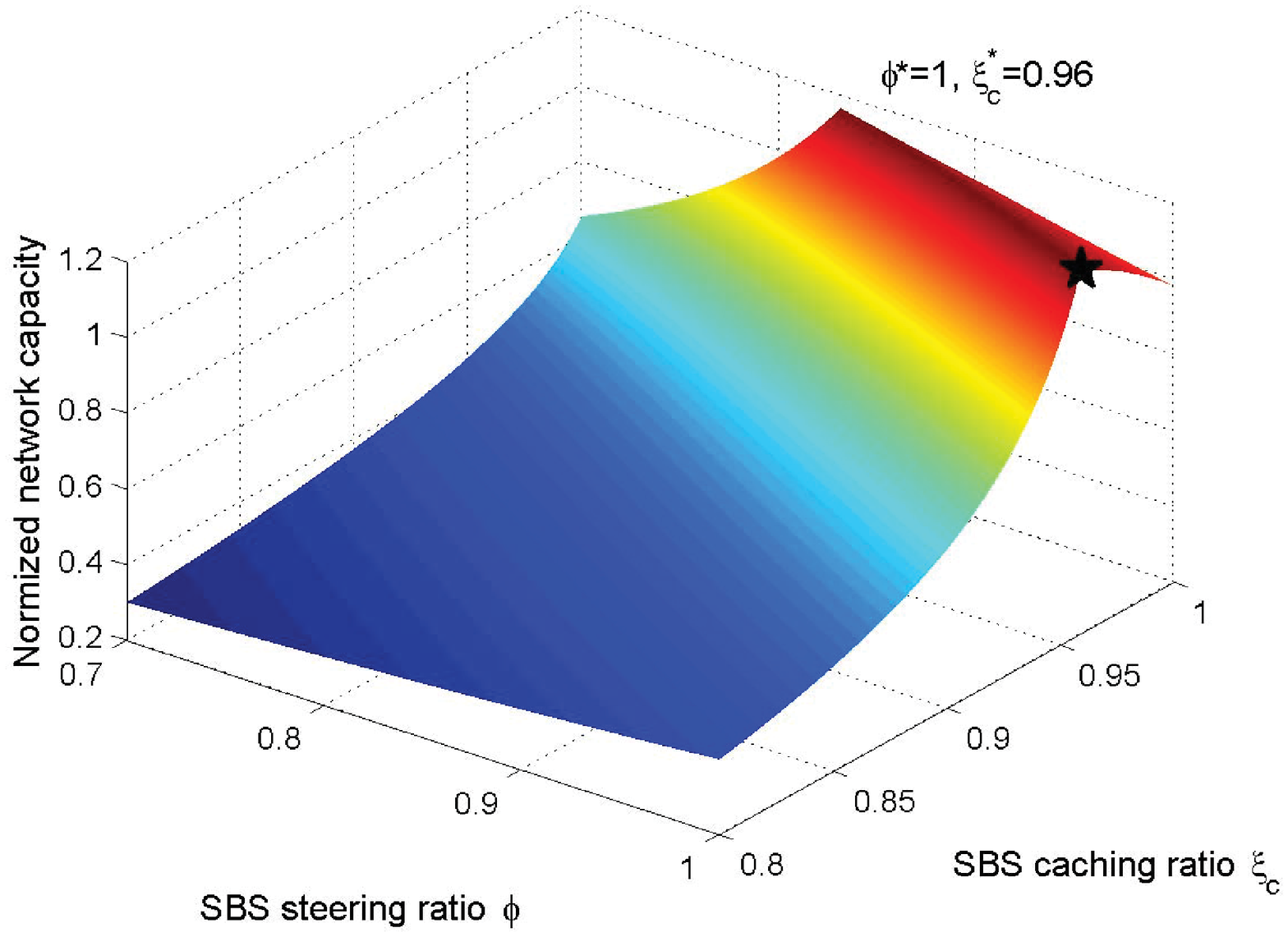}}
    	\caption{Optimal hierarchical cache.}
    	\label{fig_optimal_cache_3D}
    \end{figure*}
    
    \subsection{Optimal Hierarchical Caching}
    
    To validate the theoretical analysis, Fig.~\ref{fig_optimal_cache_3D} shows network capacity with respect to different cache deployment and traffic steering ratios, where the analytical results obtained by Propositions~2-4 are marked as the star points.
    The per user rates for radio access and backhaul are set as $\check{R}_\mathrm{RAN}$=5 Mbps and $\check{R}_\mathrm{BH}$=50 Mbps, respectively.
    $\xi_\mathrm{c}$ is the ratio of cache budget deployed at SBSs, i.e., $\xi_\mathrm{c} = C_\mathrm{s} \rho_\mathrm{s}/C$.
    According to Eq.~(\ref{eq_C_min_max}), $C_\mathrm{min}$=870 files/km$^2$, $C_\mathrm{max}$=930 files/km$^2$.
    Thus, the three subfigures correspond to the cases of Propositions~2-4, respectively.
    In Fig.~\ref{fig_optimal_cache_3D}(a), the star point is shown to achieve the maximal network capacity, validating the analysis of Proposition~2.
    In Figs.~\ref{fig_optimal_cache_3D}(b) and (c), there are multiple solutions that can achieve the maximal network capacity, including the star points.
    Furthermore, the star points also minimize the SBS cache size (i.e., minimal $\xi_\mathrm{c}$) among all the capacity-optimal schemes, indicating high content hit rate.
    When $C$=900 files/km$^2$, the SBS backhaul will become the bottleneck when $\xi_\mathrm{c}$ is lower than $\xi_\mathrm{c}^*$, degrading network capacity as shown in Fig.~\ref{fig_optimal_cache_3D}(b).
    Instead, in Fig.~\ref{fig_optimal_cache_3D}(c), the performance bottleneck is due to the MBS radio access, and thus the network capacity will decrease when $\xi_\mathrm{c}$ is lower than $\xi_\mathrm{c}^*$.
    The numerical results of Figs.~\ref{fig_optimal_cache_3D}(a)-(c) are consistent with Propositions~2-4, validating the theoretical analysis.
    
    \begin{figure}[!t]
    	\centering
    	\includegraphics[width=2.5in]{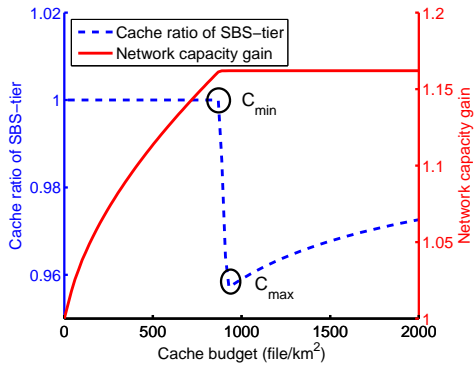}
    	\caption{Optimal cache spliting with respect to cache budget.}
    	\label{fig_optimal_planning_cache_budget}
    \end{figure}
    
    Fig.~\ref{fig_optimal_planning_cache_budget} further demonstrates the relationship between the cache budget and the optimal cache deployment, obtained by exhaustive search.
    As shown by the dash line, the optimal cache deployment can be divided into three cases.
    Firstly, all cache budget should be allocated to the SBS tier when the cache budget is insufficient, i.e., $C<C_\mathrm{min}$.
    As the cache budget achieves $C_\mathrm{min}$ and is lower than $C_\mathrm{max}$, it is shown that the ratio of cache budget allocated to the SBS tier begins to decrease.
    Furthermore, the ratio of cache budget allocated to the SBS tier increases again when the cache budget exceeds $C_\mathrm{max}$.
    %	Thus, the dash line in Fig.~\ref{fig_optimal_planning_cache_budget} illustrates Propositions 2-4.
    
    %%
    \subsection{Cache-Backhaul Trading}
    
    \begin{figure}[!t]
    	\centering
    	\includegraphics[width=2.5in]{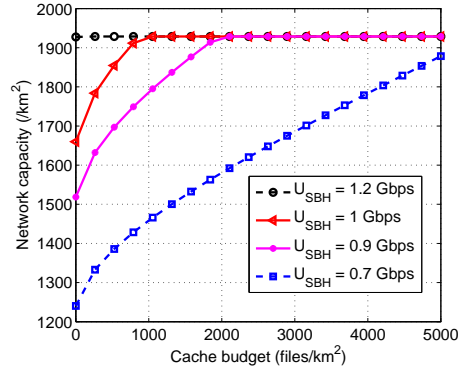}
    	\caption{Cache-enabled network capacity.}
    	\label{fig_capacity_cache_backhaul}
    \end{figure}
    
    \begin{figure}[!t]
    	\centering
    	\includegraphics[width=2.5in]{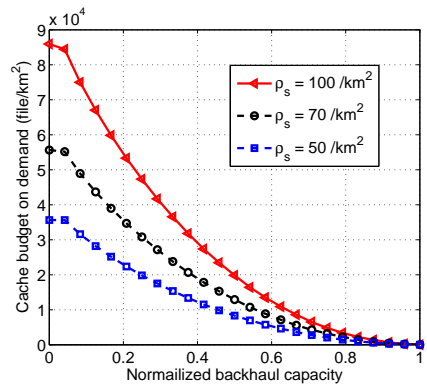}
    	\caption{Cache budget demand on different backhaul capacity.}
    	\label{fig_cache_backhaul_trading}
    \end{figure}
    
    The solid line of Fig.~\ref{fig_optimal_planning_cache_budget} presents the network capacity with respect to cache budget, which is normalized by the capacity without cache.
    As the cache budget increases, the network capacity firstly increases and then levels off as a constant.
    The reason is that the SBS backhaul is no longer the bottleneck when the cache budget achieves $C_\mathrm{min}$, and the network performance is constrained by the radio resources.	
    Fig.~\ref{fig_capacity_cache_backhaul} further illustrates the cache-enabled network capacity gain under different SBS backhaul capacities.
    Similarly, the network capacities firstly increase and then level off, and the turning points $C_\mathrm{min}$ depend on cache budgets.
    Furthermore, a larger backhaul capacity results in a smaller turning point, as shown in Fig.~\ref{fig_capacity_cache_backhaul}.
    Specifically, no cache budget is needed when the backhaul capacity is $U_\mathrm{SBH}=1.2$ Gbps, since such backhaul capacity is sufficiently large compared with SBS radio access resources.
    
    Fig.~\ref{fig_capacity_cache_backhaul} reveals the trading relationship between backhaul capacity and cache budget demands.
    Specifically, networks with insufficient backhaul capacity can be compensated by deploying cache, and the backhaul deficiency determines the amount of cache budget needed.
    The relationship between required cache budget and backhaul capacity is illustrated in Fig.~\ref{fig_cache_backhaul_trading}, where the backhaul capacity is normalized by the capacity of radio access.
    It is shown that less cache budget is needed as the backhaul capacity increases, and denser networks demand higher cache budget.
    The trading relationship between backhaul capacity and cache budget demand can be applied to cost-effective network deployment, which determines the optimal combination of backhaul capacity and cache budget.

    \subsection{Case Studies on Cost-Effective Network Deployment}
    
    \begin{figure*}[!t]
    	\centering
    	\subfloat[Cache-enabled small cells] {\includegraphics[width=2.5in]{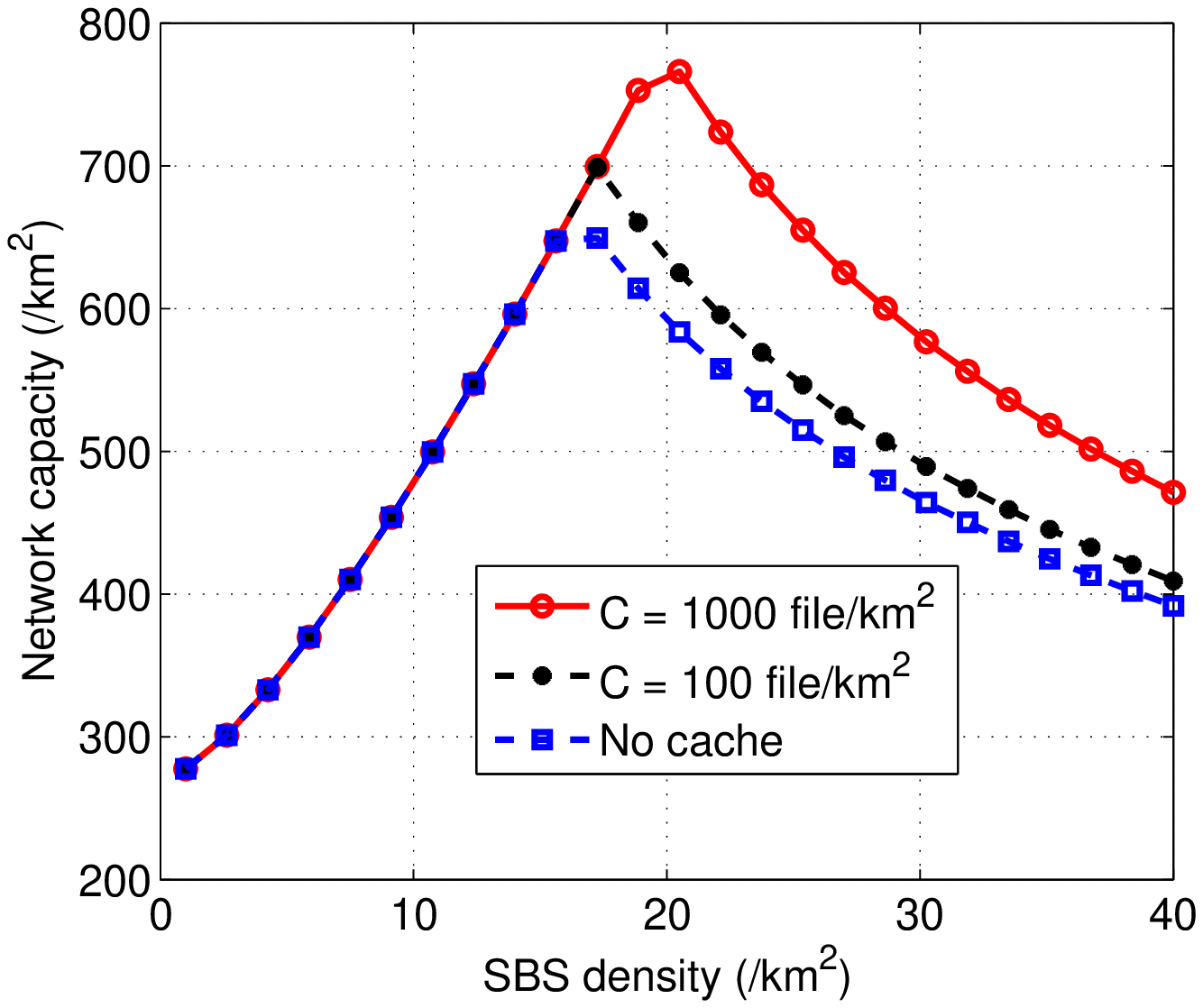}}
    	\hfil
    	\subfloat[Caching station]{\includegraphics[width=2.5in]{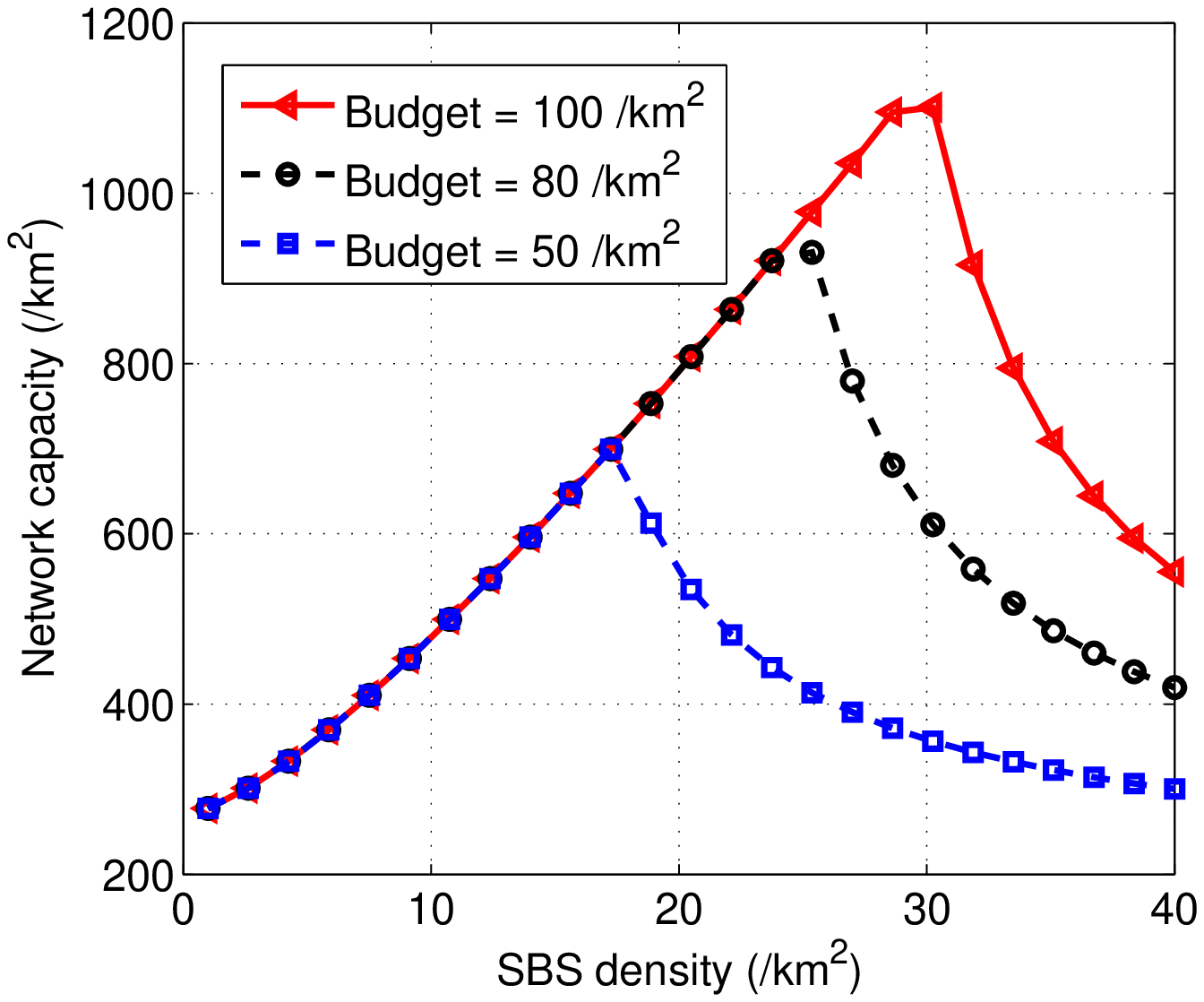}}
    	\caption{Cost-effective network deployment.}
    	\label{fig_network_deployment}
    \end{figure*}
    
    Finally, we provide case studies on cost-effective network deployment, by applying the results of  Fig.~\ref{fig_cache_backhaul_trading}.
    Fig.~\ref{fig_network_deployment}(a) illustrates the cost-optimal SBS deployment for the given backhaul deployment cost, under different caching budget $C$.
    The backhaul deployment cost is considered to increase with SBS density as well as backhaul capacity: $\rho_\mathrm{s}(1+K_\mathrm{BH} U_\mathrm{SBH}^{\zeta_\mathrm{BH}})$\footnote{$K_\mathrm{BH}$ denotes the ratio of backhaul deployment cost to the SBS equipment cost, and $\zeta_\mathrm{BH}$ reflects how the backhaul deployment cost scales with capacity.}.
    For illustration, $K_\mathrm{BH}=0.001$ and $\zeta_\mathrm{BH}=0.5$\footnote{With this setting, the backhaul deployment cost is comparable to the SBS equipment cost when backhaul capacity is 1 Gbps.}.
    When budget of the backhaul cost is 100 /km$^2$, the network capacity with respect to different SBS density is shown as Fig.~\ref{fig_network_deployment}(a).
    The network capacity is demonstrated to firstly increase and then decrease with SBS densities, falling into two regions.
    {{On one hand, the capacity of radio access increases with SBS density.
    		On the other hand, the backhaul capacity per SBS decreases with SBS density due to the constrained deployment cost.
    		Accordingly, the performance of SBS tier is constrained by the radio access resources when the SBS density is low, and becomes backhaul-constrained when the SBS density exceeds some threshold.
    		In fact, the optimal SBS density achieves the best match between radio and backhaul resource settings.}}
    
    Fig.~\ref{fig_network_deployment}(b) further demonstrates the cost-optimal deployment of caching stations, which is a special case when the SBSs have no backhaul and all content miss users are served by MBSs.
    The deployment cost is considered to increase with SBS density as well as SBS cache size, i.e., $\rho_\mathrm{s} (1+K_\mathrm{c} {C_\mathrm{s}}^{\zeta_\mathrm{C}})$\footnote{$K_\mathrm{c}$ denotes the ratio of storage cost to the cost of other modules, and $\zeta_\mathrm{C}$ reflects how storage cost scales with cache size.}.
    $K_\mathrm{C}$ and $\zeta_\mathrm{C}$ are system parameters, set as $K_\mathrm{C}=0.1$ and $\zeta_\mathrm{\textsc{C}}=0.5$ for illustration\footnote{With this setting, the storage cost is comparable to the other modules when cache size is 100 files, i.e., 10\% of all contents}.
    The network capacity with respect to SBS density is demonstrated in Fig.~\ref{fig_network_deployment}(b).
    {{The cost-optimal SBS density reflects the tradeoff between radio access capacity and content hit rate.
    		The capacity for radio access increases with SBS density, whereas the cache size decreases due to the deployment budget.
    		With low density, the SBSs are overloaded due to the constrained radio resources.
    		However, the high-dense SBSs can only serve few users due to low content hit rate, causing radio resource underutilized.
    		Thus, the optimal SBS density should balance the capacity of SBS radio access and content hit rate, to maximize network capacity.
    		The results of Fig.~\ref{fig_network_deployment} can provide insightful design criteria for cost-effective network deployment.
    	}}
    
%%%%%%%%%%%%%%%%%%%%%%%%%%%%%%%%%%%%%%%%%%%%%%%%%%%%%%%%%%%%%%%%%%%%%%%%%%%%%%%%%%%%%%%%%%%%%%%%%%%
\section{Conclusions and Future Work}
    \label{sec_conclusions}

In this paper, the cost-effective cache deployment problem has been investigated for a large-scale two-tier HetNet, aiming at maximizing network capacity while meeting file transmission rate requirements.
By conducting stochastic geometry analysis, the capacity-optimal cache sizes have been derived, which is threshold-based with respect to cache budget under the SBS-backhaul-constrained case.
The analytical results of cache budget threshold have been obtained, which characterize the backhaul deficiency and the cache-backhaul trading relationship.
{{The proposed cache deployment schemes can be applied to practical network upgrades as well as capacity enhancement.
		When the existing networks upgrade with storage units for edge caching, the optimal cache sizes of different BSs can be directly determined with the obtained cache budget threshold, based on system parameters such as base station density, radio resources, backhaul capacity, and content popularity. 
		When more cache-enabled MBSs or SBSs are deployed for capacity enhancement, the proposed method can be applied to determine the optimal cache sizes and simplify the optimization of other system parameters.}}
%The proposed cache deployment scheme can play an important role in cost-effective network deployment, by matching and fully utilizing cache, radio and backhaul resources.
{{For future work, we will optimize cache deployment based on cooperative caching scheme, where multiple SBSs can cooperate to serve users.}}

%%%%%%%%%%%%%%%%%%%%%%%%%%%%%%%%%%%%%%%%%%%%%%%%%%%%%%%%%%%%%%%%%%%%%%%%%%%%%%%%%%%%%%%%%%%%%%%%%%%

\appendices{}

\section{Proof of Lemma~1}
\label{appendix_rate_MR}
The average transmission rate of MBS radio access is given by:
%		\begin{equation}
%			\label{eq_rate_MR_1}
%			\begin{split}
%				& \mathds{E}[R_\mathrm{MR}] = \underset{ \{ N_\mathrm{MR}, d_\mathrm{m} \}} {\mathds{E}} [\frac{W_\mathrm{m}}{1+N_\mathrm{MR}}\log_2(1+\gamma_\mathrm{m})] \\
%				& = W_\mathrm{m} \underset{ \{ N_\mathrm{MR} \}} {\mathds{E}} [\frac{1}{1+N_\mathrm{MR}}] \underset{ \{ d_\mathrm{m} \}} {\mathds{E}} [\log_2(1+\gamma_\mathrm{m})].
%			\end{split}
%		\end{equation}
\begin{equation}
\label{eq_rate_MR_1}
\begin{split}				
& \mathds{E}[R_\mathrm{MR}] = W_\mathrm{m} {\mathds{E}} \left[\frac{1}{1+N_\mathrm{MR}}\right] {\mathds{E}} \left[\log_2(1+\gamma_\mathrm{m})\right] \\ & 
=  \frac{W_\mathrm{m}}{\ln 2} {\mathds{E}} \left[\frac{1}{1+N_\mathrm{MR}}\right] {\mathds{E}} [\ln(1+\gamma_\mathrm{m})]
\end{split}
\end{equation}
As $N_\mathrm{MR}$ follows Poisson distribution of mean $\lambda_\mathrm{MR}/\rho_\mathrm{m}$,
\begin{equation}
\label{eq_N_MR_aver_appendix}
{\mathds{E}} \left[\frac{1}{1+N_\mathrm{MR}}\right] = \frac{ \rho_\mathrm{m}}{\lambda_\mathrm{MR}} \left( 1-e^{-\frac{\lambda_\mathrm{MR}}{\rho_\mathrm{m}}} \right),
\end{equation}
which can be derived in the same way as Eq.~(\ref{eq_rate_MBH}) by replacing $\lambda_\mathrm{MBH}$ by $\lambda_\mathrm{MR}$.
Furthermore,
\begin{subequations}
	\label{eq_SE_MR}
	\begin{align}
	& {\mathds{E}} \left[\ln(1+\gamma_\mathrm{m})\right] = \frac{D_\mathrm{min}^2}{D_\mathrm{m}^2} \ln \left( 1+\frac{P_\mathrm{TM} D_\mathrm{min}^{-\alpha_\mathrm{m}}}{(1+\theta_\mathrm{m})\sigma^2} \right) \nonumber \\
	& +  \int_{D_\mathrm{min}}^{D_\mathrm{m}} \frac{2d}{D_\mathrm{m}^2} \ln \left( 1+\frac{P_\mathrm{TM}d^{-\alpha_\mathrm{m}}}{(1+\theta_\mathrm{m})\sigma^2} \right)  \mbox{d} d \nonumber\\
	& \geq  \frac{D_\mathrm{min}^2}{D_\mathrm{m}^2} \ln \left(\frac{P_\mathrm{TM} D_\mathrm{min}^{-\alpha_\mathrm{m}}}{(1+\theta_\mathrm{m})\sigma^2} \right) - \frac{2 \alpha_\mathrm{m} }{D_\mathrm{m}^2}  \int\limits_{D_\mathrm{min}}\limits^{D_\mathrm{m}} d \ln d  \mbox{d} d \nonumber \\
	& + \left(1-\frac{D_\mathrm{min}^2}{D_\mathrm{m}^2}\right) \ln \left( \frac{P_\mathrm{TM}}{(1+\theta_\mathrm{m})\sigma^2} \right)  \\
	%& = \ln \left( \frac{P_\mathrm{TM}}{(1+\theta_\mathrm{m})\sigma^2} \right) -  \frac{D_\mathrm{min}^2}{D_\mathrm{m}^2} \ln(D_\mathrm{min}^{\alpha_\mathrm{m}}) \nonumber\\
	%& - \frac{\alpha_\mathrm{m}}{D_\mathrm{m}^2}\left[ d^2\left( \ln d - \frac{1}{2} \right)\mid_{D_\mathrm{min}}^{D_\mathrm{m}} \right] \nonumber\\
	& = \ln\frac{P_\mathrm{TM} D_\mathrm{m}^{-\alpha_\mathrm{m}}  }{(1+\theta_\mathrm{m})\sigma^2}+\frac{\alpha_\mathrm{m}}{2}\left(1-\frac{D_\mathrm{min}^2}{D_\mathrm{m}^2}\right), \nonumber
	\end{align}
\end{subequations}
where $\theta_\mathrm{m} \sigma^2 = \mathds{E}[I_\mathrm{m}]$, and the equality of (\ref{eq_SE_MR}a) holds when $\frac{\sigma^2}{P_\mathrm{TM}} \rightarrow 0$. 
Substituting Eqs.~(\ref{eq_N_MR_aver_appendix}) and (\ref{eq_SE_MR}) into (\ref{eq_rate_MR_1}), Lemma~1 can be proved.

\section{Proof of Lemma~2}
\label{appendix_rate_SR}

The average transmission rate of SBS radio access is given by:
%		\begin{equation}
%			\label{eq_rate_SR_1}
%			\begin{split}
%			& \mathds{E}[R_\mathrm{SR}] = \underset{ \{ A_\mathrm{s} N_\mathrm{SR}, d_\mathrm{s} \}} {\mathds{E}} [\frac{W_\mathrm{s}}{1+N_\mathrm{SR}}\log_2(1+\gamma_\mathrm{s})] \\
%			& = W_\mathrm{m} \underset{ \{ A_\mathrm{s}, N_\mathrm{SR} \}} {\mathds{E}} [\frac{1}{1+N_\mathrm{SR}}] \underset{ \{ d_\mathrm{s} \}} {\mathds{E}} [\log_2(1+\gamma_\mathrm{s})].
%			\end{split}
%		\end{equation}
\begin{equation}
\label{eq_rate_SR_1}
\mathds{E}\left[R_\mathrm{SR}\right] = \frac{W_\mathrm{m}}{\ln 2} {\mathds{E}} \left[\frac{1}{1+N_\mathrm{SR}}\right] {\mathds{E}} \left[\ln(1+\gamma_\mathrm{s})\right].
\end{equation}
Similar to Eq.~(\ref{eq_rate_SBH_derived}),
\begin{equation}
\label{eq_N_SR_aver_appendix}
% \underset{ \{ A_\mathrm{s}, N_\mathrm{SR} \}} 
{\mathds{E}} \left[\frac{1}{1+N_\mathrm{SR}}\right] = \frac{\kappa\rho_\mathrm{s}}{\lambda_\mathrm{SR}} \frac{\Gamma(\kappa-1)}{\Gamma(\kappa)} \left( 1 - \frac{1}{\left(1+\frac{\lambda_\mathrm{SR}}{\kappa \rho_\mathrm{s}}\right)^{\kappa-1}} \right).
\end{equation}
As SBSs follows PPP of density $\rho_\mathrm{s}$, the PDF of transmission distance $d_\mathrm{s}$ follows
\begin{equation}
f_{d_\mathrm{s}}(d) = \frac{\mbox{d}}{\mbox{d} d} \left( 1-e^{-\pi \rho_\mathrm{s} d^2 } \right).
\end{equation}
Thus, 
\begin{subequations}
	\label{eq_SE_SR}
	\begin{align}
	& {\mathds{E}} \left[\ln(1+\gamma_\mathrm{s})\right] = \int_{0}^{\infty} \ln \left( 1+\frac{P_\mathrm{TS}d^{-\alpha_\mathrm{s}}}{(1+\theta_\mathrm{s})\sigma^2} \right) f_{d_\mathrm{s}}(d)  \mbox{d} d \nonumber\\
	& \geq  \ln \frac{P_\mathrm{TS}}{(1+\theta_\mathrm{s})\sigma^2}  - \alpha_\mathrm{s} \int_{0}^{\infty} 2 \pi \rho_\mathrm{s} d e^{-\pi \rho_\mathrm{s} d^2} \ln d \mbox{d} d \\
	& = \ln \frac{P_\mathrm{TS}(\pi \rho_\mathrm{s})^{\frac{\alpha_\mathrm{s}}{2}}}{(1+\theta_\mathrm{s})\sigma^2}  - \alpha_\mathrm{s} \int_{0}^{\infty} e^{-x} \ln x \mbox{d} x \nonumber\\
	& = \ln \frac{P_\mathrm{TS}(\pi \rho_\mathrm{s})^{\frac{\alpha_\mathrm{s}}{2}}}{(1+\theta_\mathrm{s})\sigma^2}  + \frac{1}{2} \alpha_\mathrm{s} \gamma,
	\end{align}
\end{subequations}
where $\theta_\mathrm{s} \sigma^2 = \mathds{E}\left[I_\mathrm{s}\right]$, and the equality of (\ref{eq_SE_SR}b) holds when $\frac{\sigma^2}{P_\mathrm{TS}} \rightarrow 0$. 
Substituting Eqs.~(\ref{eq_N_SR_aver_appendix}) and (\ref{eq_SE_SR}) into (\ref{eq_rate_SR_1}), Lemma~2 can be proved.

\section{Proof of Table~\ref{tab_variation}}
\label{appendix_variation}

When $\varphi$ increases, the ratio of traffic load steered to MBS backhaul and radio access both decrease according to Eq.~(\ref{eq_capacity_part}a) and (\ref{eq_capacity_part}b), hence increasing $\mu_\mathrm{MR}$ and $\mu_\mathrm{MBH}$.
On the contrary, $\mu_\mathrm{SR}$ and $\mu_\mathrm{SBH}$ both decrease, according to Eq.~(\ref{eq_capacity_part}c) and (\ref{eq_capacity_part}d).

Suppose the SBS cache size $C_\mathrm{s}$ increases to $C'_\mathrm{s} = C_\mathrm{s} + \Delta_\mathrm{s}$, and the MBS cache size $C_\mathrm{m}$ decreases to $C'_\mathrm{m}=C_\mathrm{m} + \Delta_\mathrm{m}$, where $\rho_\mathrm{s} \Delta_\mathrm{s} + \rho_\mathrm{m} \Delta_\mathrm{m}=0$.
As $\rho_\mathrm{s}>\rho_\mathrm{m}$ in practical networks, $\Delta_\mathrm{s} + \Delta_\mathrm{m} < 0$.
Denote by $\Delta P_\mathrm{Hit}^{(\mathrm{m})}$ and $\Delta P_\mathrm{Hit}^{(\mathrm{s})}$ the corresponding variations of MBS-tier and SBS-tier content hit rates, respectively.
Apparently, $\Delta P_\mathrm{Hit}^{(\mathrm{m})}<0$ and $\Delta P_\mathrm{Hit}^{(\mathrm{s})}>0$, and $\Delta P_\mathrm{Hit}^{(\mathrm{m})} + \Delta P_\mathrm{Hit}^{(\mathrm{s})} < 0$ since $\Delta_\mathrm{s} + \Delta_\mathrm{m} < 0$.
Thus, the total content hit rate decreases.
Therefore, $\mu_\mathrm{MBH}$ and $\mu_\mathrm{SBH}$ both decrease, according to Eqs.~(\ref{eq_capacity_part}b) and (\ref{eq_capacity_part}d).
In addition, $\mu_\mathrm{SR}$ also decreases as $P_\mathrm{Hit}^{(\mathrm{s})}$ increases, according to Eq.~(\ref{eq_capacity_part}c).
For $\mu_\mathrm{MR}$,
\begin{equation}
\begin{split}
& \Delta P_\mathrm{Hit}^{(\mathrm{m})} - ( \Delta P_\mathrm{Hit}^{(\mathrm{m})} + \Delta P_\mathrm{Hit}^{(\mathrm{s})}) (1-\varphi)\\
& = -\Delta P_\mathrm{Hit}^{(\mathrm{s})} + (\Delta P_\mathrm{Hit}^{(\mathrm{m})} + \Delta P_\mathrm{Hit}^{(\mathrm{s})}) \varphi < 0
%					& \Delta P_\mathrm{Hit}^{(\mathrm{s})} + ( -\Delta P_\mathrm{Hit}^{(\mathrm{m})} - \Delta P_\mathrm{Hit}^{(\mathrm{m})}) (1-\varphi) \\
%					& = -\Delta P_\mathrm{Hit}^{(\mathrm{s})} + \varphi (\Delta P_\mathrm{Hit}^{(\mathrm{m})} + \Delta P_\mathrm{Hit}^{(\mathrm{m})}) < 0,
\end{split}
\end{equation}
and thus $\mu_\mathrm{MR}$ increases with $C_\mathrm{s}$, according to Eq.~(\ref{eq_capacity_part}a).

\section{Proof of Proposition~2}
\label{appendix_insufficient_cache}

Set $C_\mathrm{s} = C/\rho_\mathrm{s}$, $C_\mathrm{m}=0$, $\varphi=\frac{\hat{\lambda}_\mathrm{SBH}} {\hat{\lambda}_\mathrm{MR} + \hat{\lambda}_\mathrm{SBH}}$.
Accordingly, $P_\mathrm{Hit}^{(\mathrm{s})} = \sum_{f=1}^{C/\rho_\mathrm{s}} q_f$, $P_\mathrm{Hit}^{(\mathrm{m})} = 0$, and $P_\mathrm{Hit} = \sum_{f=1}^{C/\rho_\mathrm{s}} q_f$.
Substituting $P_\mathrm{Hit}^\mathrm{(m)}$ and $\varphi$ into Eqs.~(\ref{eq_capacity_part}a) and (\ref{eq_capacity_part}d), we have $\mu_\mathrm{MR} = \mu_\mathrm{SBH} = \frac{\hat{\lambda}_\mathrm{MR}+\hat{\lambda}_\mathrm{SBH}}{1-P_\mathrm{Hit}}$, and $\mu_\mathrm{SR} = \frac{\hat{\lambda}_\mathrm{SR}}{1-\frac{\hat{\lambda}_\mathrm{MR}}{\hat{\lambda}_\mathrm{MR} + \hat{\lambda}_\mathrm{SBH}} (1-P_\mathrm{Hit}) }$.	
As $C < C_\mathrm{min}$,
\begin{equation}
\begin{split}
& P_\mathrm{Hit} = \sum_{f=1}^{C/\rho_\mathrm{s}} q_f < \frac{\hat{\lambda}_\mathrm{SR} - \hat{\lambda}_\mathrm{SBH}}{ \hat{\lambda}_\mathrm{MR} + \hat{\lambda}_\mathrm{SR}} \\
\Longleftrightarrow & P_\mathrm{Hit} (\hat{\lambda}_\mathrm{MR} + \hat{\lambda}_\mathrm{SR}) < \hat{\lambda}_\mathrm{SR} - \hat{\lambda}_\mathrm{SBH} \\
\Longleftrightarrow & P_\mathrm{Hit} \hat{\lambda}_\mathrm{MR} + \hat{\lambda}_\mathrm{SBH} < (1-P_\mathrm{Hit}) \hat{\lambda}_\mathrm{SR}
\end{split}
\end{equation}	
Notice that
\begin{equation}
\begin{split}
& \frac{\hat{\lambda}_\mathrm{SR}}{1-\frac{\hat{\lambda}_\mathrm{MR}}{\hat{\lambda}_\mathrm{MR} + \hat{\lambda}_\mathrm{SBH}} (1-P_\mathrm{Hit}) } - \frac{\hat{\lambda}_\mathrm{MR}+\hat{\lambda}_\mathrm{SBH}}{1-P_\mathrm{Hit}} \\
& = \frac{\hat{\lambda}_\mathrm{MR}+\hat{\lambda}_\mathrm{SBH}}{1-P_\mathrm{Hit}} \left[  \frac{\hat{\lambda}_\mathrm{SR} (1-P_\mathrm{Hit}) }{ \hat{\lambda}_\mathrm{SBH} + P_\mathrm{Hit} \hat{\lambda}_\mathrm{MR}} -1 \right] >0.
\end{split}
\end{equation}
Thus, $\mu_\mathrm{SR} > \mu_\mathrm{MR} = \mu_\mathrm{SBH}$.
According to Table~\ref{tab_variation}, $\mu_\mathrm{MR}$ and $\mu_\mathrm{SBH}$ cannot be simultaneously improved.
Therefore, $C_\mathrm{s}^* = C/\rho_\mathrm{s}, \varphi^* = \frac{\hat{\lambda}_\mathrm{SBH}} {\hat{\lambda}_\mathrm{MR} + \hat{\lambda}_\mathrm{SBH}}$ is the optimal solution to problem (P2).

\section{Proof of Propositions~3}
\label{appendix_sufficient_cache}

Firstly, we prove that $\left[C_\mathrm{s}^*, \varphi^* \right]$ satisfying Eq.~(\ref{eq_P3_opt_sufficient_C}) is feasible to constraint (\ref{eq_P3}f) in problem (P2).
As $C \geq C_\mathrm{min} $, $C_\mathrm{s}^* = C_\mathrm{min}/\rho_\mathrm{s}$ is feasible to (\ref{eq_P3}f).
When $C_\mathrm{s}^* = C_\mathrm{min}/\rho_\mathrm{s}$, the SBS-tier content hit rate is ${P_\mathrm{Hit}^{(\mathrm{s})}}^* = \frac{\hat{\lambda}_\mathrm{SR} - \hat{\lambda}_\mathrm{SBH} }{\hat{\lambda}_\mathrm{MR} +\hat{\lambda}_\mathrm{SR}}$, according to the definition of $C_\mathrm{min}$ in Eq.~(\ref{eq_C_min_max}).
As $C < C_\mathrm{max}/\rho_\mathrm{s}$, ${P_\mathrm{Hit}^{(\mathrm{m})}}^* < \frac{\hat{\lambda}_\mathrm{MR} }{\hat{\lambda}_\mathrm{MR} +\hat{\lambda}_\mathrm{SR}}$.
Thus, ${1-P_\mathrm{Hit}}^* > \frac{\hat{\lambda}_\mathrm{SBH} }{\hat{\lambda}_\mathrm{MR} +\hat{\lambda}_\mathrm{SR}}$, and $\varphi \in (0,1)$ is feasible to (\ref{eq_P3}f) in (P2).

Then, we prove that the network capacity achieves the maximum under $\left[C_\mathrm{s}^*, \varphi^* \right]$.
Substituting Eq.~(\ref{eq_P3_opt_sufficient_C}) into (\ref{eq_capacity_part}), we have $\mu^*_\mathrm{MR} = \mu^*_\mathrm{SR} = \mu^*_\mathrm{SBH} = \hat{\lambda}_\mathrm{MR}+\hat{\lambda}_\mathrm{SR}$.
Thus, the network capacity is $\hat{\lambda}_\mathrm{MR}+\hat{\lambda}_\mathrm{SR}$.
Adding constraints (\ref{eq_P3}b) and (\ref{eq_P3}d), we can prove that the maximal network capacity cannot exceed $\hat{\lambda}_\mathrm{MR}+\hat{\lambda}_\mathrm{SR}$.
Therefore, Eq.~(\ref{eq_P3_opt_sufficient_C}) guarantees the optimality of capacity.

In addition, we prove that the network capacity is smaller than $\hat{\lambda}_\mathrm{MR} +\hat{\lambda}_\mathrm{SR}$ if $C_\mathrm{s} \leq C_\mathrm{s}^*$, by contradiction.
Assume there exist a solution $[C_\mathrm{s}', \varphi']$ with network capacity of $\hat{\lambda}_\mathrm{MR} +\hat{\lambda}_\mathrm{SR}$, where $C_\mathrm{s}'\leq C_\mathrm{min}/\rho_\mathrm{s}$.
According to Eqs.~(\ref{eq_P3_opt_sufficient_C}a) and (\ref{eq_P3_opt_sufficient_C}d), we have
\begin{subequations}
	\label{eq_app_P3_opt_proof_1}
	\begin{align}					
	& {P_\mathrm{Hit}^{(\mathrm{M})}}' + (1-P'_\mathrm{Hit}) (1 - \varphi') \leq \frac{\hat{\lambda}_\mathrm{MR}}{ \hat{\lambda}_\mathrm{MR}+ \hat{\lambda}_\mathrm{SR}}, \\
	& (1-P'_\mathrm{Hit}) \varphi' \leq \frac{\hat{\lambda}_\mathrm{SBH}}{ \hat{\lambda}_\mathrm{MR}+ \hat{\lambda}_\mathrm{SR}}.
	\end{align}
\end{subequations}
In addition
\begin{subequations}
	\label{eq_app_P3_opt_proof_2}
	\begin{align}
	& {P_\mathrm{Hit}^{(\mathrm{M})}}' + (1-P'_\mathrm{Hit}) (1 - \varphi') \nonumber \\
	%& = 1-{P_\mathrm{Hit}^{(\mathrm{s})}}' - (1-P'_\mathrm{Hit})\varphi' \nonumber \\
	& \geq 1-{P_\mathrm{Hit}^{(\mathrm{s})}}' - \frac{\hat{\lambda}_\mathrm{SBH}}{ \hat{\lambda}_\mathrm{MR}+ \hat{\lambda}_\mathrm{SR}} \\
	%& = 1-{P_\mathrm{Hit}^{(\mathrm{s})}}' - (1-P^*_\mathrm{Hit})\varphi^* \\
	& > 1 - {P_\mathrm{Hit}^{(\mathrm{s})}}^* - (1-P^*_\mathrm{Hit})\varphi^* \\
	& = \frac{\hat{\lambda}_\mathrm{MR}}{ \hat{\lambda}_\mathrm{MR}+ \hat{\lambda}_\mathrm{SR}},
	\end{align}
\end{subequations}
where (\ref{eq_app_P3_opt_proof_2}a) is based on (\ref{eq_app_P3_opt_proof_1}b), (\ref{eq_app_P3_opt_proof_2}b) is due to condition (\ref{eq_P3_opt_sufficient_C}b), (\ref{eq_app_P3_opt_proof_2}c) holds as $P_\mathrm{Hit}^{(\mathrm{s})}$ increases with $C_\mathrm{s}$, and (\ref{eq_app_P3_opt_proof_2}d) comes from (\ref{eq_P3_opt_sufficient_C}).
As (\ref{eq_app_P3_opt_proof_2}) is contradictory with (\ref{eq_app_P3_opt_proof_1}a), there exists no $C_\mathrm{s}'\leq C_\mathrm{min}/\rho_\mathrm{s}$ to achieve the maximal network capacity.
As content hit rate decreases with SBS cache size, $[C_\mathrm{s}^*,\varphi^*]$ achieves the maximal content hit rate among the capacity-optimal solutions.		
%Hence, Proposition~3 is proved.

\bibliographystyle{IEEEtran}

\end{document}